\documentstyle[preprint,prc,eqsecnum,aps,epsf]{revtex}

\tightenlines   

\begin{document}
\draft

\def\lsim{\thinspace{\hbox to 8pt{\raise -5pt\hbox{$\sim$}\hss{$<$}}}\thinspace}
\def\rsim{\thinspace{\hbox to 8pt{\raise -5pt\hbox{$\sim$}\hss{$>$}}}\thinspace}

\title{ Full-Folding Optical Potentials for Elastic Nucleon-Nucleus
Scattering based on Realistic Densities}

\author{ Ch.~Elster, S.P.~Weppner}
\address{
Institute of Nuclear and Particle Physics,  and
Department of Physics, \\ Ohio University, Athens, OH 45701}

\author{ C.R.~Chinn 
\footnote{present address: Aret\'{e} Associates, P.O. Box 16269, Arlington,
VA 22215}}
\address{
 Department of Physics and Astronomy, Vanderbilt University,
Nashville, TN  37235}


\date{\today}

\maketitle
 
\begin{abstract}

Optical model potentials for elastic nucleon nucleus scattering are
calculated for a number of target nuclides from a full-folding integral
of two different realistic target density matrices together with full
off-shell nucleon-nucleon t-matrices derived from two different Bonn
meson exchange models. Elastic proton and neutron scattering observables
calculated from these full-folding optical potentials 
are compared to those obtained from `optimum
factorized' approximations in the energy regime between 65 and 400~MeV
projectile energy. The optimum factorized form is found to provide a
good approximation to elastic scattering observables obtained from the
full-folding optical potentials, although the potentials differ somewhat
in the structure of their nonlocality.
\end{abstract}


\pagebreak


\narrowtext


\section{Introduction}

\hspace*{10mm}
Progress in the rigorous treatment of the multiple scattering of
nucleons from nuclei has led to the need to study the influence of
the full target nuclear density matrix in the scattering observables.
In first order the spectator expansion of multiple scattering 
theory requires a convolution of the fully off-shell nucleon-nucleon
(NN) scattering amplitude with the nuclear wave functions of the target.
This opens the possibility to assess the influence of the
target wave functions on elastic proton and neutron scattering
observables.

\hspace*{10mm}
In its most general form, the first order single scattering optical
potential within the framework of the spectator expansion is given by
the triangle graph shown in Fig.~1. Since there is one loop, the
evaluation of the graph requires a three-dimensional integration
involving the fully-off-shell two-nucleon scattering amplitude and a
realistic nuclear density matrix. 
Usually, one makes the assumption that the NN amplitude varies slowly as
a function of its arguments compared to the nuclear density matrix.
This corresponds to the argument that the range of the NN force is small
compared to the size of the nucleus and leads to the approximate
nonrelativistic form $t(q)\rho(q)$ for the first-order nucleon-nucleus
optical potential. Full-folding calculations, avoiding this
approximation, have already been performed by several groups based on
either the KMT approach \cite{FFC,FFO} or a g-matrix approach
\cite{hugo1,hugo2} using various models for the off-shell density matrix as
well as different models for the NN amplitudes. 
In general, this work indicates that an improved
treatment of the off-shell structure of the optical potential improves
the description of the observables.

\hspace*{10mm}  
Our approach to elastic scattering from nuclei is based on the spectator
expansion of multiple scattering theory \cite{Corr,Sicil,TT}. Here the
first order term involves two-body interactions between the projectile
and one of the target nucleons. Due to the many-body nature of the free
propagator for the projectile+target nucleon system, a theoretical
treatment of this many-body propagator as affected by the residual
target nucleus is included. The calculation of the optical potential
presented in this paper relies on two basic inputs. One is the
fully off-shell NN t-matrix, which represents the current understanding
of the nuclear force, and the other is the nuclear wave function of the
target, representing the best understanding of the ground state of the
target nucleus. To account for the modifications of the free propagator
inside the nucleus, the same mean field potentials are used from which
the ground state wave functions are derived. There are {\it no}
adjustable parameters present in this calculation. 

\hspace*{10mm}
The motivation for ongoing work on this topic is twofold. First,
elastic and inelastic nucleon-nucleus scattering provide an
important and sensitive test for theoretical corrections at the
first-order level of the optical potential ({\it e.g.} as
given by possibly genuine modifications
of the NN interaction in the nuclear environment
and off-shell effects).  Rigorous microscopic
calculations are required for discerning these effects. 
Second, a better understanding of the theoretical details of the
optical potential are needed to construct realistic and physically
sound wave functions representing continuum nucleons in the
interior of the nucleus.  These wave functions will become vital
for future theoretical needs in high-energy coincidence experiments
(at TJNAF e.g.), inelastic scattering studies, and for understanding the
reactions in heavy ion experiments
involving nuclei far from the drip lines. Here experiments involving
the scattering of exotic nuclei from single nucleon targets should 
should benefit from full-folding type calculations in order to test
the predicted density distributions  of halo-like nuclei.

\hspace*{10mm}
The structure of the paper is as follows. In Section~II we review the
relevant formalism for the single-scattering optical potential and
introduce the full-folding procedure as used in our calculations.
In Section~III we discuss the model densities employed and describe the
calculations of the full folding optical potentials. Elastic scattering
results for neutron and proton scattering form a variety of nuclei in the
energy regime between 65 and 400 MeV are discussed in Section~IV. Our
conclusions are presented in Section~V.

\section{The full-folding optical potential}

\hspace*{10mm}
The standard approach to elastic scattering of a strongly interacting 
projectile from
a target of $A$ particles is the separation of the Lippmann-Schwinger
equation for the transition amplitude
\begin{equation}
T = V + V G_0(E) T  \label{eq:2.1}
\end{equation}
into two parts, namely an integral equation for $T$:
\begin{equation}
T = U + U G_0(E) P T  , \label{eq:2.2}
\end{equation}
where $U$ is the optical potential operator and defined by a second
integral equation
\begin{equation}
U = V + V G_0(E) Q U.  \label{eq:2.3}
\end{equation}
In the above equations the operator $V$ represents the external
interactions between the projectile and the target nucleons. 
Therefore the Hamiltonian for the
$(A+1)$ particle system is given by
\begin{equation}
H=H_{0}+V .  \label{eq:2.4}
\end{equation}
The free propagator $G_{0}(E)$ for the projectile-target system is 
given by
\begin{equation}
G_{0}(E)=(E-H_{0}+i\epsilon)^{-1}.  \label{eq:2.5}
\end{equation}
The potential operator $V={\sum_{i=1}^{A}} v_{0i}$ consists of the
two-body potential $v_{0i}$ acting between the projectile and
the $i$th target nucleon. The operators $P$ and $Q$ are projection 
operators, $P+Q=1$, and $P$ is defined such that Eq.~(\ref{eq:2.2}) is
solvable. In this case, $P$ is conventionally taken to project onto 
the elastic channel, such that $[G_{0},P]=0$. The free 
Hamiltonian is given by
\begin{equation}
H_{0}=h_{0}+H_{A}   \label{eq:2.6}
\end{equation}
where $h_{0}$ is the kinetic energy operator for the projectile
and $H_{A}$ stands for the target Hamiltonian. Thus the projector $P$
can be defined as
\begin{equation}
P={|\Phi_{A}\rangle\langle\Phi_{A}|\over{\langle\Phi_{A}|\Phi_{A}
\rangle}}   \label{eq:2.7}
\end{equation}
where $|\Phi_{A}\rangle$ corresponds to the ground state of the target 
and fulfills 
\begin{equation}
H_{A}|\Phi_{A}\rangle= E_{A}|\Phi_{A}\rangle  \label{eq:2.8}
\end{equation}
With these definitions the transition operator for elastic scattering 
can be defined as ${T_{el}=PTP}$, in which case Eqs.~(\ref{eq:2.2}) 
 becomes
\begin{equation}
T_{el}=PUP + PUPG_{0}(E)T_{el}.  \label{eq:2.9}
\end{equation}

\hspace*{10mm}
The fundamental idea of the spectator expansion for the optical potential
is an ordering of the scattering process according to the number
of active target nucleons interacting directly with the projectile.
The first order term involves two-body interactions between the 
projectile and one of the target nucleons, i.e.
\begin{equation}
U = \sum_{i=1}^{A}\tau_{i} , \label{eq:2.10}
\end{equation}
where the operator $\tau_{i}$ is derived to be
\begin{eqnarray}
\tau_i &=& v_{0i} + v_{0i} G_0(E) Q \tau_i \nonumber \\
&=& v_{0i} + v_{0i}G_0(E) \tau_i - v_{0i} G_0(E) P \tau_i \label{eq:2.11} \\
&=& \hat{\tau_i} - \hat{\tau_i} G_0(E) P \tau_i . \nonumber
\end{eqnarray}
For elastic scattering only $P\tau_{i}P$, or equivalently
$\langle\Phi_A| \tau_i | \Phi_A\rangle$ need to be considered,
\begin{equation}
\langle\Phi_A| \tau_i | \Phi_A\rangle = \langle\Phi_A| \hat{\tau_i}|
 \Phi_A\rangle - \langle\Phi_A| \hat{\tau_i}| \Phi_A\rangle \frac {1}
 {(E-E_A) - h_0 + i\varepsilon} \langle\Phi_A| \tau_i | \Phi_A\rangle ,
 \label{eq:2.12}
\end{equation}
where $\hat{\tau_i}$ is defined as the solution of
\begin{equation}
\hat{\tau_i} = v_{0i} + v_{0i} G_0(E) \hat{\tau_i}. \label{eq:2.13}
\end{equation}
It should be noted that Eqs.~(\ref{eq:2.3}) to (\ref{eq:2.13}) all
follow in a straightforward  derivation and 
correspond
to the first order Watson scattering expansion \cite{Watson}. If the
projectile$\,-\,$target nucleon interaction is assumed to be the
same for all target nucleons and if isospin effects are neglected
then the KMT scattering integral equation \cite{KMT} can be 
directly derived from
the first order Watson scattering expansion.

\hspace*{10mm}
Since Eq.~(\ref{eq:2.12}) is a simple one-body integral equation,
the principal problem is to find a solution of Eq.~(\ref{eq:2.13}),
which has a  many-body character due to
$G_0(E)=(E -h_{0} -H_{A} +i\varepsilon)^{-1}$.
If the propagator $G_{0}(E)$
is expanded in the the spirit of the spectator
expansion within a single particle description, 
one obtains in first order \cite{med2,med1}
\begin{equation}
G_i(E) = [(E-E^i) -h_0 -h_i -W_i + i\varepsilon]^{-1}, \label{eq:2.14}
\end{equation}
where $h_i$ is
the kinetic energy of the $i$th target particle and
$W_i=\sum_{j\neq i}v_{ij}$. 
The quantity $W_i$ represents the force acting between the struck
target nucleon and the residual (A-1) nucleus.
Then the operator
$ \hat{\tau_i}$  of Eq.(\ref{eq:2.13}) can be written as
\begin{eqnarray}
\hat{\tau_i}&=&v_{0i} + v_{0i} G_i(E)\hat{\tau_i} \nonumber \\
  &=& t_{0i} + t_{0i} g_i W_i G_i(E) \hat{\tau_i}.   \label{eq:2.15}
\end{eqnarray}
Here the operators $t_{0i}$ and $g_i$ are defined as
\begin{equation}
t_{0i} = v_{0i} + v_{0i} g_i t_{0i} \label{eq:2.17}
\end{equation}
and
\begin{equation}
g_i =[ (E-E^i) - h_0 -h_i + i\varepsilon]^{-1}. \label{eq:2.18}
\end{equation}
The operator $t_{0i}$ can be identified with the free NN t-matrix,
and in lowest order the operator $\hat{\tau_i}$ of Eq.~(\ref{eq:2.15})
is given by $\hat{\tau_i} \approx t_{0i}$. From now on we consider
for clarity in  presentation only this case.

\hspace*{10mm}
The matrix element
$\langle\Phi_A|\tau_i | \Phi_A\rangle$ given in
Eq.~(\ref{eq:2.12})  represents the full-folding optical potential and
is given explicitly as
\begin{equation}
\langle{\bf k}' | U |{\bf k}\rangle = 
\langle{\bf k}'\Phi_{A} | \sum_{\alpha=p,n} {\tau_{\alpha}}
|{\bf k}\Phi_{A}\rangle , \label{eq:2.19} 
\end{equation}
where $\alpha$ represents the sum over the target protons and neutrons.
Since $\langle{\bf k}' | U |{\bf k}\rangle$ is the solution of the sum
of the
one-body integral equations represented by Eq.~(\ref{eq:2.12}), 
it is sufficient to consider the driving term
\begin{equation}
\langle{\bf k}' |\hat{U}|{\bf k}\rangle = 
\langle{\bf k}'\Phi_{A} | \sum_{\alpha=p,n} \hat{\tau}_{\alpha}
|{\bf k}\Phi_{A}\rangle , \label{eq:2.20}
\end{equation}
where $\hat{\tau}_{\alpha}$ is given by Eq.~(\ref{eq:2.13}).

\hspace*{10mm}
Inserting a complete set of momenta for the struck target nucleon 
before and after the collision Eq.~({\ref{eq:2.20}}) reads
\begin{eqnarray}
\hat{U}\left({\bf k},{\bf k^{\prime}}\right)=
\sum_{\alpha=p,n} \int d^{3}{\bf p^{\prime}}
d^{3}{\bf p} \left\langle {{\bf k^{\prime}}{\bf p^{\prime}}}
\mid \hat{\tau}_{\alpha} (\epsilon)
\mid {{\bf k p}} \right\rangle \rho_{\alpha} \left({{\bf p^{\prime}
+\frac{{\bf k'}}{A}}}, {\bf p}+\frac{{\bf k}}{A}\right)
\delta^{3} ({\bf k^{\prime}} + {\bf p^{\prime}} -
{\bf k} - {\bf p}).  \label{eq:2.21}
\end{eqnarray}
The momenta ${\bf k'}$ and ${\bf k}$ are the final and initial momenta
of the projectile in the frame of zero total nucleon-nucleus 
momentum. The structure of Eq.~({\ref{eq:2.21}}) is represented
graphically by Fig~1, which also illustrates the momenta ${\bf p'}$ and
${\bf p}$. The proton and neutron density matrices are given
by $\rho_{\alpha}$. 
 By evaluating the $\delta$-function and introducing the
variables 
${\bf q}={\bf k'}-{\bf k}$, 
${\bf K}={1\over{2}}({{\bf k} + {\bf k^{\prime}}})$ and
${\bf \hat{p}}={1\over{2}}({{{\bf p^{\prime}}}+{{\bf p}}})$  we obtain
\begin{eqnarray}
\hat{U}({\bf q},{\bf K})=\sum_{\alpha=p,n}
\int d^{3}{\bf\hat{p}}\;\left\langle {\bf k'},
{\bf\hat{p}}-{1\over{2}}{{\bf q}}\left|\;\hat{\tau}_{\alpha}
(\epsilon)\;\right|{{\bf k}},{\bf\hat{p}}+{1\over{2}}{{\bf q}}\right\rangle\;
  \nonumber \\
\;\;\;\;\;\;\;\;\;\;\;\;\;\;\;\;\;\;
\rho_{\alpha}\left({{\bf \hat{p}}}-{{A-1}\over{A}}{{{\bf q\over{2}}}}+
{{\bf K}\over{A}},
{{\bf \hat{p}}}+{{A-1}\over{A}}{{{{\bf q}}\over{2}}+{{{{\bf K}}}\over{A}}
}\right) . \label{eq:2.22}
\end{eqnarray}
A change of the integration variable from $\bf\hat{p}$ to ${\bf P}={\bf\hat p}
+ {{{{\bf K}}}\over{A}}$, accounting for the recoil of the nucleus, gives
\cite{pttw}
\begin{eqnarray} 
\hat{U}({\bf q},{\bf K})=\sum_{\alpha=p,n}
\int d^{3}{\bf P}\;\left\langle {\bf k'},{\bf P}-{{\bf q}\over{2}}-
{{\bf K}\over{A}}\left|\;\hat{\tau}_{\alpha}                
(\epsilon)\;\right|{\bf k},{\bf P}+{{\bf q}\over{2}}-
{{\bf K}\over{A}}\right\rangle\; \nonumber \\
\;\;\;\;\;\;\;\;\;\;\;\;\;\;\;
\rho_{\alpha}\left({\bf P}-{{A-1}\over{A}}{{{\bf q}\over{2}}},
{\bf P}+{{A-1}\over{A}}{{\bf q}\over{2}}\right).  \label{eq:2.23}
\end{eqnarray}  
The NN amplitude ${\hat \tau}_{\alpha}$ 
in Eq.~({\ref{eq:2.23}}) is evaluated in the zero momentum
frame of the nucleon-nucleus system. The relationship to the corresponding
matrix element evaluated in the zero momentum frame of the two nucleons
is given by
\begin{eqnarray}
{\left\langle {\bf {k'}},{\bf P}-{{\bf q}\over{2}}-
{{\bf K}\over{A}}\left|\;\hat{\tau}_{\alpha}                
(\epsilon)\;\right|{\bf {k}},{\bf P}+{{\bf q}\over{2}}-
{{\bf K}\over{A}}\right\rangle}_{NA} = 
\eta ({\bf P},{\bf q},{\bf K})
{\left\langle {\cal K}',-{\cal K}'|\;\hat{\tau}_{\alpha}(\epsilon)\;
|{\cal K},-{\cal K}\right\rangle}_{NN},
\end{eqnarray}
where  ${\bf\cal{K}'}=\frac {1}{2}({\bf k'}-({\bf P}-\frac{{\bf q}}{2}-
\frac{{\bf K}}{A}))$ and ${\bf\cal{K}}=\frac {1}{2}({\bf k}-
({\bf P}+{{\bf q}\over{2}}- {{\bf K}\over{A}}))$
are the nonrelativistic final and initial nuclear momenta in the 
zero momentum frame of the NN system.
The factor $\eta({\bf P},{\bf q},{\bf K})$ is the M\o ller factor for the 
frame transformation \cite{joachain} and is given by
\begin{equation}
\eta({\bf P},{\bf q},{\bf K})=\left[{{E_N({\cal K}')E_N(-{\cal K}')
E_N({\cal K})E_N(-{\cal K})}\over{E_N(k')E_N({\bf P}-{{\bf q}\over{2}}-
{{\bf K}\over{A}})E_N(k)E_N({\bf P}+{{\bf q}\over{2}}-    
{{\bf K}\over{A}})}}\right], \label{eq:2.25}
\end{equation}
where $E_N({\bf k})$
is the relativistic kinetic energy of a nucleon of
momentum $k$. This factor imposes the Lorentz invariance of the flux.
With this frame transformation taken into account, the full-folding
optical potential of Eq.~({\ref{eq:2.23}}) can be written as
\begin{eqnarray} 
\hat{U}({\bf q},{\bf K})=\sum_{\alpha=p,n}
\int d^{3}{\bf P}\;
\eta({\bf P},{\bf q},{\bf K})\;
\hat{\tau}_{\alpha}\left({\bf q},{1\over{2}}({{A+1}\over{A}}{\bf K}-{\bf P});
\epsilon\right)\; \nonumber \\
\;\;\;\;\;\;\;\;\;\;\;\;\;\;\;
\rho_{\alpha}\left({\bf P}-{{A-1}\over{A}}{{{\bf q}\over{2}}},
{\bf P}+{{A-1}\over{A}}{{\bf q}\over{2}}\right). \label{eq:2.26}
\end{eqnarray} 
Here the arguments of ${\hat \tau}_{\alpha}$ are ${\bf q}= {\bf k}'
-{\bf k} = {\bf\cal{K}'}-{\bf\cal{K}}$ and $\frac {1}{2}({\bf\cal{K}'}+
{\bf\cal{K}})= \frac {1}{2}(\frac {A+1}{A} {\bf K} -{\bf P})$.

\hspace*{10mm}
The two-nucleon amplitude $\hat{\tau}_{\alpha}$ is calculated from the
free NN t-matrix according to Eqs.~(\ref{eq:2.15}) and (\ref{eq:2.17}) 
at an appropriate energy $\epsilon$. In principle, this energy should be the
beam energy minus the kinetic energy of the center of mass of the interacting
pair less the binding energy of the struck particle.
Following this argument, $\epsilon$ should be coupled to the integration
variable ${\bf P}$. The full-folding calculations of
Refs.~\cite{hugo1,hugo2} are carried out along this vain. For our
calculations we take a different approach, namely we fix $\epsilon$
at the two-body center-of-mass (c.m.) energy corresponding to free
NN scattering at the beam energy so that the same laboratory energy
applies to the two-body and nuclear scattering. 
This approach has been applied in earlier work \cite{FFO} and is also
used in the work of the Surrey Group \cite{FFC}.

\section{Models for the Off-Shell Density}

\hspace*{10mm}
The evaluation  of the full-folding optical potential as given in 
Eq.~({\ref{eq:2.26}}) requires a nuclear density matrix, which 
in a single particle picture is given as
\begin{equation}
\rho_\alpha ({\bf \tilde p'},{\bf \tilde p}) = \sum_i 
\Psi_{\alpha,i}^{\dagger} ({\bf \tilde p'}) \Psi_{\alpha,i}
({\bf \tilde p}) \label{eq:3.1}
\end{equation}
Here $\Psi_{\alpha,i}({\bf \tilde p})$ are the wave functions 
describing the  single particle nuclear ground state. The index $\alpha$
stands for protons and neutrons, respectively,
and the total nuclear ground state
is given by the sum of the two.
In order to achieve consistency with our
formulation of incorporating effects of the `nuclear medium'
on the scattering process we choose as model density matrices the
ones from which the nuclear mean fields $W_i$ are derived (see
e.g. Eq.~({\ref{eq:2.15}})). The models used are a
non-relativistic reduction of a Dirac-Hartree calculation \cite{DH} and a
non-relativistic
Hartree-Fock-Bogolyubov (HFB) structure calculation \cite {HFB,Gogny}.
The Dirac-Hartree calculation is a spherical solution of the
one-body Dirac equation assuming a scalar potential
and the time component of a
vector potential field. The nonrelativistic HFB microscopic nuclear
structure calculation uses the parameterized effective finite-range,
density dependent Gogny D1S effective 
NN interaction. The parameter of the  Gogny
D1S interaction are fitted to a certain set of stable nuclei. For this
case an axial harmonic oscillator basis is used.

\hspace*{10mm}
The details of the Dirac-Hartree (DH) calculation 
leading to the density matrices employed in our calculations are
given below.
The wave function $\Psi_i({\bf r})$ is a solution of the one-body
Dirac-Hartree equation  and given by \cite{TIMORA}
\begin {equation}
{\Psi}_{\beta}({\bf r}) \equiv {\Psi}_{n,\nu,m,t_z}({\bf r})
= \left({\begin{array}{cc}
i{[\frac{G_{t_z,n,\nu}(r)}{{r}}}]\;{\bf \phi}_{\nu m} \\
-{[\frac{F_{t_z,n,\nu}(r)}{{r}}}]\;{\bf \phi}_{-\nu m}\;
\end{array}}
\right) \zeta_{t_z} \label{eq:3.3}
\end {equation}
Here $t_z$ stands for the z component of the isospin and $n$
for the principal quantum number. The phase convention is 
taken from Ref.~\cite{TIMORA}. During this derivation we prefer
to omit the index $\alpha$.
The spherical harmonics are determined by ${\bf\phi}_{\nu m}$ which is
defined as
\begin{equation}
{\bf \phi}_{\nu m}=\sum_{m_l,m_s}
\langle lm_l{1\over{2}}m_s|
l{1\over{2}}jm\rangle Y_{l}^{m_l}(\Omega )\chi_{m_s}, \label{eq:3.3b}
\end{equation}
where $Y_{l}^{m_l}(\Omega)$ is a spherical harmonic and $\chi_{m_s}$
a Pauli spinor.
The quantum number $\nu$ uniquely defines $j$ and $l$ as
\begin{equation}
j = |\nu|-\frac{1}{2}\;\;\;\;, l =\left\{\begin{array}{cc}
\nu,\;\; \nu>0 \\
-(\nu+1),\;\;\nu<0\end{array}\right\}. \label{3.3b}
\end{equation}
We used the code {\it Timora} \cite{TIMORA} 
and the parameter sets given therein to
generate the functions $G$ and $F$ given in Eq.({\ref{eq:3.3}) for
the nuclei studied in this paper.

Under the assumption of orthogonal single particle states the
density matrix is given in coordinate space by
\begin{eqnarray}
 & &\rho ({\bf r}',{\bf r})  =  \sum_{n\nu m} \Psi^{\dagger}_{n\nu m t_z}
 ({\bf r}')  \Psi_{n\nu m t_z} ({\bf r}) \nonumber \\
&=& \sum_{n\nu}\left[ \frac {G_{t_z,n,\nu}(r')}{r'}
\frac {G_{t_z,n,\nu}(r)}{r} \sum_m \phi_{\nu m}(r') \phi_{\nu m}(r)
+ \frac {F_{t_z,n,\nu}(r')}{r'}\frac {F_{t_z,n,\nu}(r)}{r}
 \sum_m \phi_{-\nu m}(r') \phi_{-\nu m}(r) \right]. \label{eq:3.04}
\end{eqnarray}
Here we should point out that in order to obtain a 
density matrix which we can apply in our 
formulation of the optical potential, we have a {\bf 1}-operator
between the Dirac wave functions $\Psi_{n\nu m t_z}$, and then
treat $\rho ({\bf r}',{\bf r})$ as nonrelativistic density matrix.
The orthogonality of the spin states leads to 
$\delta_{m_s,m_s'}$ and thus to $m_l=m_{l'}$.
Taking advantage of the symmetry properties of the Clebsch-Gordon 
coefficients leads to
\begin{equation}
\rho ({\bf r}',{\bf r})  =  \sum_{n\nu} \left[ \frac
{G_{t_z,n,\nu}(r')}{r'} \frac {G_{t_z,n,\nu}(r)}{r}
+ \frac {F_{t_z,n,\nu}(r')}{r'}\frac {F_{t_z,n,\nu}(r)}{r}
 \right] \frac {2j+1}{2l+1} \sum_{m_l} Y_l^{* m_l}(r')
Y_l^{m_l}(r). \label{eq:3.05}
\end{equation}

\vspace{2mm}
\hspace*{10mm}
The calculation of the full-folding  optical potential
${\hat U}({\bf q},{\bf K})$ requires the nuclear density
matrix in momentum space. Thus we need to
double Fourier transform $\rho ({\bf r}',{\bf r})$
to obtain the density $\rho({\bf \tilde p'},{\bf \tilde p})$  
in the rest frame of the nucleus. This frame is characterized
 by  the momenta $\bf\tilde{p}$ and $\bf\tilde{p}'$ and the density
matrix is obtained by
\begin{equation}
\rho_{\alpha}({\bf \tilde{p}'},{\bf \tilde{p}}) = 
\frac{1}{8\pi^3} \int d^3{\bf r}^{\prime}
 e^{-i {\bf r}^{\prime}\cdot{\bf \tilde{p}}
^{\prime}} \int d^3{\bf r} e^{i {\bf r}\cdot{\bf\tilde{p}}} 
\rho_\alpha ({\bf r}',{\bf r}), \label{eq:3.4}
\end{equation}
where we again indicate with the index $\alpha$ that we have to
obtain the density matrix for protons as well as neutrons. 
Using the standard expansion of a plane wave, the angular integration
in Eq.~(\ref{eq:3.4}) can be easily carried out, and we obtain
for the density matrix
\begin{eqnarray}
\rho_{\alpha}({{\bf \tilde{p}'}},{{\bf \tilde{p}}}) &=& \frac{1}{2\pi^2}
\sum_J (2J+1) \sum_l 
P_l(\cos \theta_{\tilde{p},\tilde{p}'})    \nonumber \\
& [  &  \int dr' r' \; j_l(\tilde{p}'r')
F_{\alpha,t_z,n,\nu}(r') \;\int dr\; r\; j_l(\tilde{p}r) 
F_{\alpha,t_z,n,\nu}(r) + \nonumber \\ 
& & \int dr' r' \; j_l(\tilde{p}'r') G_{\alpha,t_z,n,\nu}(r')
\int dr\; r\; j_l(\tilde{p}r)
G_{\alpha,t_z,n,\nu}(r) \;\; ].  {\label{eq:3.7}}
\end {eqnarray}

\hspace*{10mm}
The density matrix $\rho_{\alpha}({\bf \tilde{p}'},{\bf \tilde{p}})$ 
given in Eqs.~(\ref{eq:3.4}) or (\ref{eq:3.7})
is defined in the rest frame of the nucleus. In order to apply 
$\rho_{\alpha}({\bf \tilde{p}'},{\bf \tilde{p}})$ in our calculation 
of the full-folding 
optical potential for nucleon-nucleus scattering, we have to evaluate
the function at the corresponding momenta in the nucleon-nucleus frame.
This is facilitated by variable transformations ${\bf p}=
{\bf \tilde{p}-{k\over{A}}}$ and ${\bf {p}'}=
{\bf \tilde{p}'-{k'\over{A}}}$, which takes into account recoil. As
an aside, not including recoil would mean the transformation
${\bf p'}= {\bf \tilde{p}'-{k\over{A}}}$.

\hspace*{10mm}
For the calculation of the density matrices derived from a
non-relativistic Hartree-Fock Bogolyubov (HFB) calculation based on the
Gogny-D1S NN interaction we employ essentially the same procedure as
described above. The wave functions are created in r space by a code
provided by Berger \cite{HFB} and are represented in a axially 
symmetric harmonic oscillator basis. A double Fourier transform is then 
performed using the oscillator
basis and summing over all harmonic oscillator quantum numbers. 
This choice of basis takes
advantage of the fact that the Fourier transform of a harmonic
oscillator is again a harmonic oscillator. The density matrix is
given by
\begin{equation}
\rho_{\alpha} ({\bf \tilde p'},{\bf \tilde p}) = \sum_{i,i'}
\rho^{i,i'} \varphi_{i'}^{\dagger} ({\bf \tilde p}') \varphi_i
({\bf \tilde p}), \label{eq:3.8}
\end{equation}
where the indices $i,i'$ count the harmonic oscillator basis states and
$\rho^{i,i'}$ is the density matrix in the oscillator basis.
Again, the index $\alpha$ distinguishes between protons and neutrons.
The basis states are explicitly given by
\begin{equation}
\varphi_i({\bf \tilde p})= \sum_{m}  A_{i,m_l}(\beta,\gamma)
\;e^{{-\beta  p_z}^2} \;H_i ({\beta,{ \tilde p_z}})
\;\;e^{{-\gamma  p_r}^2} \;L_i^{|m|} ({{\gamma, \tilde p_r}})
e^{im\theta}.
 \label{eq:3.9}
\end{equation}
Here ${\tilde p_z}$ is the projection of the momentum along the z-axis
and $p_r$  the radial momentum, $\beta,\gamma$ are 
harmonic oscillator constants.
$H_{i}(\beta,{\tilde p_z})$ are the Hermite polynomials and 
$L_i^{|m|}(\gamma,{\tilde p_r})$ the Laguerre polynomials.
The size of the harmonic oscillator basis used depends on the size of the
nucleus, {\it e.g.} the size of the basis for $^{16}$O is 12 shells
whereas for $^{90}$Zr it is 16 shells. 
It should be noted that the indices $i$ and $i'$
are not independent. The size of the basis sets needed makes the
calculation of $\rho_{\alpha}({\bf \tilde p'},{\bf \tilde p})$ quite lengthy,
especially for heavier nuclei. 

\hspace*{10mm}
In order  to calculate the optical potential $\hat U ({\bf q},{\bf K})$
as given in Eq.~(\ref{eq:2.26}), we need the density matrix as function
of the momentum transfer ${\bf q}$ and
${\bf P = \hat p + \frac{K}{A}}$ as indicated in Eq.~(\ref{eq:2.23}).
In these variables 
the density matrix is related to the density
profile $\rho_\alpha(q)$ of the nucleus by
\begin{equation}
\rho_\alpha(q)=\int d^3{\bf P} \rho_{\alpha}\left(
{\bf P}-{{A-1}\over{A}}{{{\bf q}\over{2}}},
{\bf P}+{{A-1}\over{A}}{{\bf q}\over{2}}\right). \label{eq:3.10}
\end{equation}
The normalization is chosen such that $\rho_\alpha(q=0)=Z$ or $N$, the number
of protons or neutrons, respectively.

\hspace*{10mm}
In practice we used the relation given in Eq.~({\ref{eq:3.10}})
for testing our numerical integration schemes with the simple
harmonic oscillator density given in Ref. {\cite{FFO}}. In order
to determine how well the two model density matrices 
presented here describe the experimentally determined proton
distribution, we calculate the proton density profiles $\rho_p(q)$ for
both the DH and the HFB models for each nucleus we consider.
In the following we want to discuss two cases, namely $^{16}$O
and $^{90}$Zr. In Fig.~2 we compare the density profiles
calculated from the DH and HFB models to the experimental
proton distribution{\cite{edata}}. Overall the DH profile follows
the experimental distribution closer than the HFB profile.
The HFB profile is shifted to larger momenta indicating that the
HFB model slightly underpredicts the radius of the proton
distribution of $^{16}$O. This feature will be visible in the 
proton-scattering observables for $^{16}$O calculated with
the HFB model. In the close-up of the minimum of the density
profile it can be seen that both model densities slightly 
deviate from the experimental profile.
In Fig.~3 we carry out the same comparison for a heavier nucleus,
$^{90}$Zr. Here both model densities follow the experimental proton
distribution {\cite{edata}} very closely. The close-up of the minimum reveals
that the HFB profile deviates only slightly from the experimental profile.
This is a general trend, the heavier nuclei are better described by the
model profiles. In fact, the proton distribution of
$^{16}$O represents the worst case of disagreement of the model profiles
with the experimental profiles. This is understandable since the HFB
model is known to provide a better representation of the larger nuclei.
 
\section{Results and Discussion}

\subsection{Details of the Calculation}
\hspace*{10mm}
In this paper the study of the elastic scattering of
neutrons and protons from spin-zero target nuclei at energies that 
range from 65 to 400 MeV (incident projectile energy) is strictly
first order in the spectator expansion. Here the connection
to the propagator $G_0(E)$ due to the coupling of the initially
struck target nucleon  to the residual target is considered to be
first-order. The full-folding optical potential is calculated as
outlined in Section II, specifically as given in Eq.~({\ref{eq:2.26}}),
using the model densities described in Section III. 
The calculations for scattering at energies smaller than 200~MeV
take into account the coupling of the struck target nucleon to the
residual nucleus via the mean field potential $W_i$,
which is chosen to be consistent with the model density employed.
Details of this procedure are given in Refs.~\cite{med2,med1}.
Calculations using the Dirac-Hartree densities (and the corresponding
potential $W_i$) are labeled DH, while 
those using the Hartree-Fock-Bogolyubov
densities (and corresponding potentials $W_i$) are labeled HFB.

\hspace*{10mm}
The convolution of the fully off-shell density matrix $\rho_\alpha$
with the fully off-shell NN t-matrix, and the M\o ller frame 
transformation factor $\eta({\bf P},{\bf q},{\bf K})$ 
as given in Eq.~({\ref{eq:2.25}})
is carried out in three dimensions without partial wave
decomposition and the integration is performed using Monte Carlo
integration techniques.
Our algorithm uses Quasi-Random numbers {\cite{sobol}},
together with
importance sampling, which according to our tests has the advantage
of needing significantly 
fewer integration points than algorithms based on 
conventional `random number' generators or Gauss-Legendre integration
to obtain the same accuracy. Quasi random numbers provide a `uniform'
random distribution over the integration space.

\hspace*{10mm}
Aside from the density matrices, the fully off-shell NN t-matrix is
another crucial ingredient in the calculation of ${\hat U}({\bf q},{\bf K})$.
The calculations presented use the free
NN interaction based upon the full Bonn potential {\cite{Bonn}}.
This interaction includes the effects of
relativistic kinematics, retarded meson propagators as given by 
time-ordered perturbation theory, and iterative and crossed meson-exchanges
with $NN$, $N\Delta$, and $\Delta\Delta$ intermediate states.
For the calculations involving projectile energies  greater then 300 MeV we
employ an extension of the Bonn model above pion-production threshold
\cite{D52}.
In this model pion production is described through the decay of the
delta isobar with a width obtained consistently from the imaginary part
of the one-pion loop diagram for the delta self-energy.
It is also  to be understood that we perform all spin summations in obtaining
${\hat U}({\bf q},{\bf K})$. This reduces the required NN t-matrix elements
to a spin independent component (corresponding to the Wolfenstein amplitude
A) and a spin-orbit component (corresponding to the Wolfenstein amplitude
C). Since we are assuming that we have spin saturated nuclei,
the components of the NN t-matrix depending  on the
spin of the struck nucleon vanish.
For the proton nucleus scattering calculations the Coulomb interaction 
between
the projectile and the target is included using the exact formulation 
described in Ref. {\cite{coul}}.

\hspace*{10mm}
A common approximation to the full-folding expression of Eq.~({\ref{eq:2.26}}),
which still preserves the non-local character of the NN t-matrix, is obtained
as follows. If 
one observes that the nuclear size is significantly larger than the range
of the NN interaction, the amplitude $\hat{\tau}_\alpha$ is expected to be the 
slower varying quantity in the integral of Eq.~({\ref{eq:2.26}}).
This argues for the method of optimum factorization \cite{pttw,ernst}
which proceeds via an
expansion of $\hat{\tau}_\alpha$ (including the factor $\eta({\bf q,K,P})$)
in ${\bf P}$ about a fixed value ${\bf P_0}$.
The reference momentum ${\bf P_0}$ is determined by requiring that the 
contribution of the first derivative term be minimized. In the elastic
scattering case this contribution can be made to vanish if ${\bf P_0}$
is chosen to be zero. For further details we refer to  Ref {\cite{pttw}}.
After the integration over the density matrix to produce the 
diagonal density profile $\rho_\alpha(q)$ (Eq.~({\ref{eq:3.10}}) the 
`optimum factorized' or `off-shell $t\rho$' form of the optical potential
is given by
\begin{equation}
{\hat U}_{fac}({\bf q},{\bf K})=\sum_{\alpha=p,n}
\;\eta({\bf q},{\bf K})\;\hat{\tau}_\alpha
\left({\bf q},{{A+1}\over{2A}}{\bf K}
,\epsilon\right)\; \rho_{\alpha}\left(q \right) \label{eq:4.1}
\end{equation}
Here the non-local character of the optical potential is solely 
determined by the off-shell NN t-matrix.
For harmonic oscillator model densities it has been shown
for light nuclei that the optimum factorized form 
represents the non-local character of ${\hat U}({\bf q},{\bf K})$
qualitatively {\cite{FFC,FFO}} when applied within the KMT formalism 
to first order at intermediate energies.
When comparing  elastic scattering observables obtained from full-folding
optical potentials to those obtained
from `off-shell $t\rho$' optical potentials, the scope is two-fold.
First, we employ here realistic models of the nuclear density for light as
well as heavy nuclei. Second we extend this comparison toward energies
below 100 MeV where it could be expected that the nucleon-nucleus scattering
calculation samples the optical potential further off-shell and thus 
the optimum factorized form may not be as good an approximation.

\subsection{Elastic Scattering Results}

\hspace*{10mm}
Elastic scattering calculations from several spherical nuclei 
are carried out at a variety of energies between 65 and 400 MeV
to allow for comparisons between 
results obtained from the full-folding optical potentials with
those arising from the factorized off-shell `$t\rho$' approximation.

\hspace*{10mm}
The scattering observables for elastic proton scattering from 
$^{16}$O are displayed in Fig.~4. The solid line represents the 
calculation with the full-folding optical potential based on the DH
density and the Bonn t-matrix defined above pion-production threshold,
while the dashed line represents the optimum factorized form
as defined in Eq.~({\ref{eq:4.1}}).  Both calculations  are
based on the free NN t-matrix.
Since the two calculations give very similar results,
it can be concluded that the bulk of the non-locality of the optical
potential, which affects the elastic scattering observables, 
must come from the off-shell structure of the NN t-matrix.
The off-shell structure of the nuclear density matrix plays an
insignificant  role
for elastic scattering observables at these high energies. A similar
conclusion was already drawn in Ref. \cite{FFO} and is here confirmed
using realistic densities. In order to illustrate the effect of
the different density profiles for the DH and the HFB models 
(as shown in Fig.~2) on the elastic observables, we display 
two calculations based on the factorized optical potential for the DH
(solid line) and the HFB model (dashed line) in Fig.~5. As already discussed
in Section III, especially in the case of $^{16}$O, the HFB density
profile is shifted to larger momenta compared to the DH profile.
This translates directly into a slight shift of the first minimum of
the differential cross-section to larger angles and a slightly smaller
angular spacing of the diffraction minima. We carried out similar comparisons
of full-folding and optimum factorized optical potentials 
for heavier nuclei, but there the disagreement between the density
profiles of the two models is much smaller then for $^{16}$O
and consequently the prediction of the observables are very similar.

\hspace*{10mm}
Another effect worthwhile to study in this context is the influence of
the M\o ller factor \cite{joachain}, which takes into account the 
transformation of the NN t-matrix evaluated in the NN c.m. frame to the
zero momentum frame of the nucleon-nucleus system. This frame transformation
can be viewed as a relativistic effect and its importance should increase
with higher scattering energies. For these reasons we want to consider
it's effect on the elastic scattering observables for proton scattering
from $^{16}$O at 500 MeV (Fig.~6). The M\o ller factor
$\eta({\bf P},{\bf q},{\bf K})$ as given in Eq.~({\ref{eq:2.25}})
is a function of three vector momenta and is part of the full-folding
integral of Eq.~(\ref{eq:2.26}). The solid line in Fig.~6 represents
the calculation of $\hat{U}({\bf q},{\bf K})$ as given in
Eq.~(\ref{eq:2.26}). In the spirit of the optimum factorized 
approximation $\eta({\bf P},{\bf q},{\bf K})$ can be expanded around
a fixed value ${\bf P_0}$ (here ${\bf P_0}$=0), thus becoming 
independent of the integration variable ${\bf P}$.
This expansion corresponds to considering $\eta({\bf q},{\bf K})$
at a fixed angle between ${\bf q}$ and ${\bf K}$, specifically here
$\Theta=90^o$. The dashed line therefore in Fig.~6 corresponds
to evaluating the optical potential according to
\begin{eqnarray}
\hat{U}({\bf q},{\bf K})=\sum_{\alpha=p,n}
\eta({\bf q},{\bf K})_{\Theta=90^o}\;
\int d^{3}{\bf P}\;
\hat{\tau}_{\alpha}\left({\bf q},{1\over{2}}({{A+1}\over{A}}{\bf K}-{\bf P});\epsilon\right)\; \nonumber \\
\;\;\;\;\;\;\;\;\;\;\;\;\;\;\;
\rho_{\alpha}\left({\bf P}-{{A-1}\over{A}}{{{\bf q}\over{2}}},
{\bf P}+{{A-1}\over{A}}{{\bf q}\over{2}}\right) \label{eq:4.2}
\end{eqnarray}
The dashed and solid lines in Fig.~6 are almost indistinguishable.
This infers the conclusion that $\eta({\bf q},{\bf K})_{\Theta =90^o}$
is a very good representation of the exact expression given in 
Eq.~(\ref{eq:2.25}). In order to illustrate the total effect due to the
inclusion of the M\o ller factor, the dotted line in Fig.~6 represents
a calculation with $\eta({\bf q},{\bf K})$ set to one in Eq.~(\ref{eq:4.2}).

\hspace*{10mm}
At energies below 200 MeV, calculations of elastic observables not only
incorporate the effects of the nuclear structure models within the 
full-folding procedure but also via the mean field force ( given by the
structure model) which couples the struck target nucleon to the residual
nucleus. Thus it is hoped that  the influence of different structure
models on the elastic observables is observable.
In Fig.~7 we display the elastic observables for proton scattering from
$^{40}$Ca at 100 MeV laboratory energy employing the DH model for the
density as well as the mean field force $W_i$.
In Fig.~8 the corresponding calculation is done
using the HFB model. In both figures the solid line represents the
full-folding calculation, and the dashed line the factorized
off-shell `$t\rho$' approximation. All calculations 
contain the modification due to the mean field $W_i$.
The off-shell structure of the nuclear density matrix in the
full-folding procedure has at this lower energy a slightly larger
effect on the spin observables then at higher energies. In addition,
the angular distribution of the differential cross section diffracts
at slightly larger angles 
in the full-folding calculations compared to those based on the
factorized form. This trend is also be observed for the heavier nuclei
$^{90}$Zr and $^{208}$Pb (Figs.~9 and 10).

\hspace*{10mm}
The elastic observables for proton scattering from $^{90}$Zr at 
80 MeV are displayed in Fig.~9. Here the difference between the two
model densities employed is almost negligible for $d\sigma\over{d\Omega }$
and $A_y$. Only for the spin rotation function the difference given by 
using two different structure models is at higher angles as large
as the effect of using the factorized approximation. This result,
namely that the observables predicted by the two different models
under consideration are so similar, is not surprising in
that both models predict an almost identical density profile
(Fig.~3). The effect of the off-shell structure of the nuclear
density matrix is relatively small as the comparison between the
full-folding (solid) line and corresponding factorized (dashed) calculation
shows. In the case of proton scattering from $^{208}$Pb at 65 MeV
(Fig.~10) a comparison between a full-folding calculation and the 
factorized approximation reveals the same trends as the observables
for the lighter nuclei. At large angles the full-folding calculation
falls below the one given by the factorized form, and in this case also
below the data. The inclusion of the off-shell structure of the nuclear
density matrix makes the nucleus appear slightly larger, which
becomes apparent in the shifted diffraction pattern of
$d\sigma\over{d\Omega }$.

\hspace*{10mm}
It is difficult to extract properties of nonlocal potentials
from elastic scattering observables. Nonlocal effects are presumably
more important in inelastic processes which depend on the
nucleon-nucleus interaction such as quasielastic electron scattering
reactions.
In order to gain more insight
into the difference between a full-folding optical potential
and the factorized off-shell `$t\rho$' approximation to this 
potential, we plot in Figs. 11 ($^{40}$Ca) and 14 ($^{208}$Pb)
the real and imaginary parts of the on-shell value of 
$\hat{U}({\bf q},{\bf K})$ as a function of the orbital angular
momentum L.
We separate the cases $J=L+{1\over{2}}$ and $J=L-{1\over{2}}$ to 
isolate the effect of the spin-orbit force. As is seen in both figures,
the full-folding (solid lines) and the factorized (dashed lines)
on-shell values of the imaginary parts of the optical potential are
quite similar. In both cases the real part of the on-shell
values of the optical potentials exhibit an increasing suppression
for smaller L as the nonlocal effects of the density matrix as well
as the NN t-matrix are treated more adequately in the full-folding
procedure. It should be emphasized again that Figs. 11 and 14 only
show the value of $\hat{U}({\bf q},{\bf K})$ fulfilling the
on-shell condition ${\bf q}\cdot{\bf K}=0$ and ${\bf q}^2 +4{\bf K}^2
= 4 {\bf k_0}^2$ with ${\bf k_0}$ being the on-shell relative momentum for 
proton-nucleus (NA)
scattering and do not display any off-shell behavior inherent in the
potentials. After iteration to obtain the Watson optical
potential (Eq.~\ref{eq:2.11}) and then in the integral equation 
(Eq.~\ref{eq:2.9}), the on-shell elements of the NA t-matrix display
much smaller differences between the full-folding  and it's factorized,
on-shell $`t\rho$' approximation. For $^{208}$Pb the differences in
the real parts are nearly insignificant (Fig.~15), whereas for $^{40}$Ca 
a slight suppression of the real part for small L remains  for the
full-folding calculation compared to the factorized approximation.
However, the differences occur mainly for smaller L where the imaginary
part of the potentials as well as the t-matrices are relatively large.
Because the absorption is significant for these low partial waves, 
the elastic observables are not particularly sensitive to the differences
in the real parts as displayed in Figs. 13 and 16.

\hspace*{10mm}
In Fig. 17 total neutron cross section data for $^{12}$C, $^{16}$O,
$^{28}$Si, $^{40}$Ca, $^{90}$Zr, and $^{208}$Pb are shown along
with various calculations of $\sigma_{tot}(E)$ at a number of 
energies. Because the data are so extensive, the `usual' procedure
has been reversed in the plotting of these cross sections so that 
the data are represented by dotted curves, and the discrete points
correspond to calculated results. The solid diamonds represent the
full-folding calculations as described in Section II. All calculations
are based on a DH model for the nuclear density. For energies
$\leq 200$ MeV the modification of the free propagator through the
DH mean field is included as described in Ref.~\cite{med1}. It has 
been shown in Ref.\cite{med2} that for higher energies this 
modification of the free propagator becomes negligible. The open
circles represent calculations based on the factorized, off-shell
$`t\rho$' form using the same NN t-matrix.  A general trend to
be observed in Fig. 17 is the slightly lower value of $\sigma_{tot}(E)$
obtained from a full-folding calculation compared to the factorized
approximation. This trend is almost independent of the energy and
the nucleus under consideration and is consistent with the observation
that full-folding calculations of the differential cross sections 
fell slightly below the values given by a factorized calculation.

\hspace*{10mm}
At this point it is worthwhile to investigate whether the interactions
of the projectile with the target nucleus is uniformly distributed
over the entire nucleus or if specific regions of the nucleus
play a more dominant role in the scattering process at intermediate
energies. For our study we chose neutron scattering at 200 MeV
projectile energy and consider contributions from specific shells
to the total cross sections. We employ the DH model for the density
and remove outer shells of protons as well as neutrons.
Then we recalculate the scattering from the remaining `inner core',
which is chosen to be doubly magic, so that it is bound.
The results of this procedure for $^{16}$O, $^{40}$Ca, and 
$^{208}$Pb are given in Table I. As a technical detail, when 
calculating the scattering in these tests we treated the 
targets as being infinitely heavy to exclude recoil effects, which 
would be larger for smaller cores. In order to give an estimate of
the size of the recoil effect on the total cross section we give the
values of $\sigma_{tot}$ calculated with and without recoil in Table I.
The values for the total cross section for neutron scattering
for `inner cores' of 100, 40, 16, and 4 nucleons are given as 
entries of the corresponding nuclei. The entries in Table I marked
`n.b.' indicate that e.g. in the case of $^{208}$Pb the DH calculation
with only 8 neutrons and 8 protons did not result in a bound
system using the parameters of $^{208}$Pb given for those 16 nuclei.
The calculated rms radii for the `inner cores' under consideration
for $^{16}$O, $^{40}$Ca, and  $^{208}$Pb are listed in  Table~II.
This table also contains the rms radii for the proton and neutron
distributions for the above mentioned nuclei as given by the DH model.
Columns 3 and 4 of Table II compare the percentage of the volume filled
by the `inner core' if either the corresponding rms radius is used
(column 3) or the radius is taken to be proportional to $A^{1\over{3}}$
(column 4). The percentage Of the calculated total cross section
contribution from the inner core nucleus is given in column 5.
The numbers suggest that the nucleons in the interior of the
nucleus contribute to the total cross sections with a percentage
slightly larger then the volume they occupy when the volume
is based on the crude estimate $r\sim A^{1\over{3}}.$ This leads
to the conclusion that all nucleons in the nucleus almost equally
contribute to the scattering process. We performed a similar
study at 100 MeV and 500 MeV projectile energy and did not find any
significant deviations from the ratios ${\sigma_{tot}(core)/{\sigma_{tot}}}$
as given in Table II at 200 MeV.

\section{Conclusion}

\hspace*{10mm}
We have calculated the full-folding integral for the first-order
optical potential within the framework of the spectator expansion
of multiple scattering theory. These optical potentials are based
on realistic models for the nuclear density matrix, namely a
Dirac-Hartree and a Hartree-Fock-Bogolyubov model along with the
full Bonn meson exchange model for the NN t-matrix. Recoil and 
frame transformation factors are implemented in the calculation in their
complete form.
We calculated elastic scattering observables for a variety of light
and heavy nuclei at projectile energies from 65 to $\sim$400~MeV
laboratory energy. At energies below 200 MeV we included the
modification of the free propagator due to the coupling of the
struck target nucleon to the residual nucleus via the same mean
 field used to model the effect of the nuclear medium is
intrinsically consistent with the nuclear structure. The
predictions from these rigorous calculations of elastic nucleon 
nucleus observables provide excellent agreement with the experimental 
data in this energy regime.

\hspace*{10mm}
We tested the validity of the factorized off-shell `$t\rho$' 
approximation in the energy regime between 65 and 400 MeV and found
that this approximation, which only retains the non-locality 
given through the NN t-matrix, is even at lower energies a very
good representation of the full-folding calculation as far as the
elastic nucleon-nucleus observables are concerned. Differences between
the factorized approximation and the full calculation of the optical
potential are present predominantly in lower partial waves. However
due to the cumulative effect of many partial waves the elastic 
observables do not reflect these differences.
It should be noted that in {\it e.g.} inelastic scattering  of nucleons
from nuclei or quasielastic electron scattering those differences
between full-folding calculations and the corresponding factorized
approximation may become more significant.
We also studied the contribution of the interior structure of the nucleus
to the total cross section and find that all nucleons in the nucleus
contribute almost uniformally to the scattering process.

\vfill
\acknowledgments
The authors want to express their gratitude and appreciation
towards R.M. Thaler for the many stimulating, helpful and critical 
discussions during the major stages of this project.

This work was performed in part under the auspices of the U.~S.
Department of Energy under contracts No. DE-FG02-93ER40756 with 
Ohio University, DE-AC05-84OR21400 with Martin Marietta Energy 
Systems, Inc., and DE-FG05-87ER40376 with Vanderbilt University.  
We thank the Arctic Region Supercomputing Center (ARSC) and the Ohio
Supercomputer Center (OSC) for the use of their facilities. These
resources were made available through the metacenter regional
alliance
project funded by the Advanced Scientific Computing Program of the
National Science Foundation, Award number ASC-9418357 and Grant 
No.~PHS206 from OSC. We also thank the Pittsburgh Supercomputer Center
(PSC) for the use of their facilities under Grant No. PHY950010P
as well as the National Energy Research Supercomputer Center
(NERSC) for the use of their facilities 
under the FY1996 Massively Parallel Processing Access Program.




\newpage

\begin{table}
\caption{Total cross sections for neutron scattering from 
$^{16}$O, $^{40}$Ca, and $^{208}Pb$ as well as from inner shells of
those nuclei. The entries printed in boldface are the ones for which
the rms-radii and ratios are calculated in Table II.}
\begin{tabular}{|c||c|c|c|c|c|c|}

&$\sigma_{tot}[b]$ & $\sigma_{tot}[b]$ & $\sigma_{tot}[b]$
& $\sigma_{tot}[b]$ &
$\sigma_{tot}[b]$&  $\sigma_{tot}[b]$\\
&                & no recoil      & core of 4 & core of 16 &core of 40 &
core of 100 \\ \hline
$^{16}$O & .423   & .419   &{\bf .120}& .419 & & \\ \hline
$^{40}$Ca& .925   & .921   & n.b. &{\bf .419} & .921& \\ \hline
$^{208}$Pb& 3.38  & 3.37   & n.b. & n.b. &{\bf .960}& 2.01 \\
\end{tabular}
\end{table}

\vspace{15mm}

\begin{table}
\caption{Rms-radii for the proton and neutron distributions of 
$^{16}$O, $^{40}$Ca, and $^{208}Pb$ as well as from inner shells (cores).
The last three columns give the  ratios of the volumes of the cores to
the total nucleus as well as the ratios of the calculated total neutron
cross sections. The numbers used to determine the latter are the ones
printed in bold in Table I.}
\begin{tabular}{|c||c|c||c|c||c|}
  & rms-radius(full)[fm]  &rms-radius(core)[fm] & & &  \\
&(proton,neutron)&[core]:(proton,neutron)&${\langle{rms_{core}}^3\rangle\over
{\langle {rms}^3\rangle}}$ & ${A_{core}\over{A}}$
&${\sigma_{tot}(core)\over{\sigma_{tot}}}$ \\ \hline
$^{16}$O  & (2.63, 2.60)  & [4]:(1.96, 1.95)
&  42\% & 25\%     &  29\% \\ \hline
$^{40}$Ca& (3.39, 3.33)  & [16]:(2.63, 2.60)
&   48\%  & 40\%     &  45\% \\ \hline
$^{208}$Pb&(5.40, 5.67)  & [40]:(3.72, 4.91)
&  47\%  &19\%    &  29\%  \\
\end{tabular}
\end{table}

\newpage

\noindent
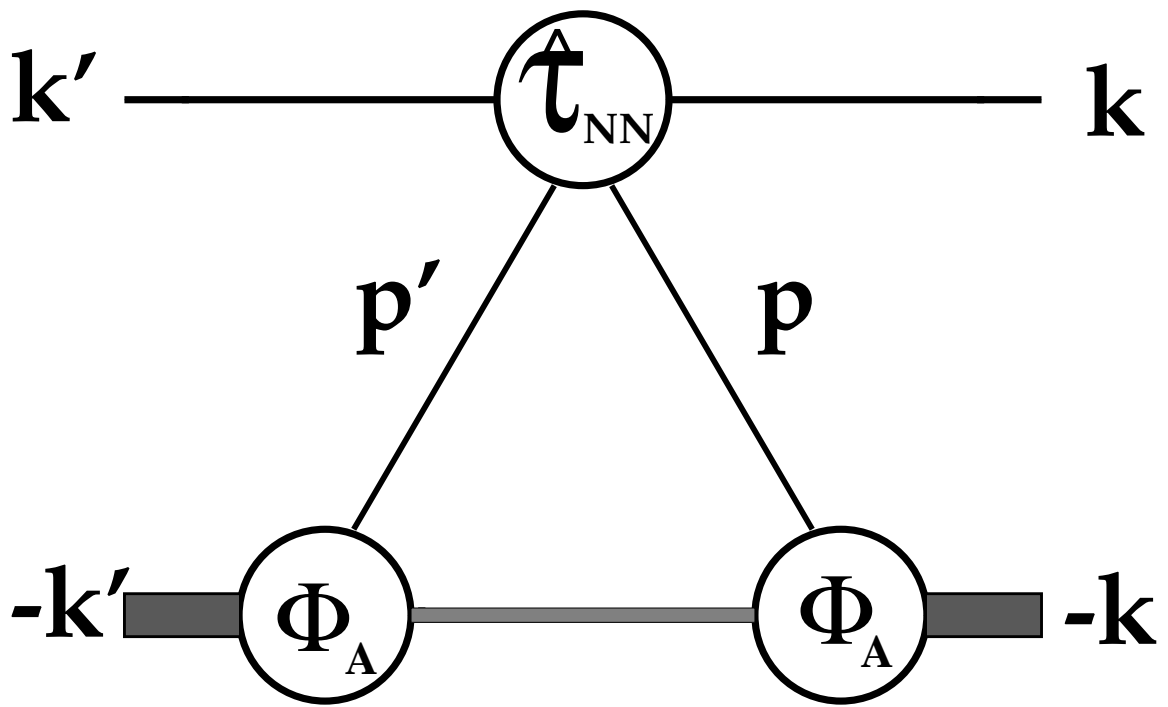
\begin{figure}
\caption{ Diagram for the optical potential matrix element for  the
single-scattering term. \label{fig1}}
\end{figure}

\noindent
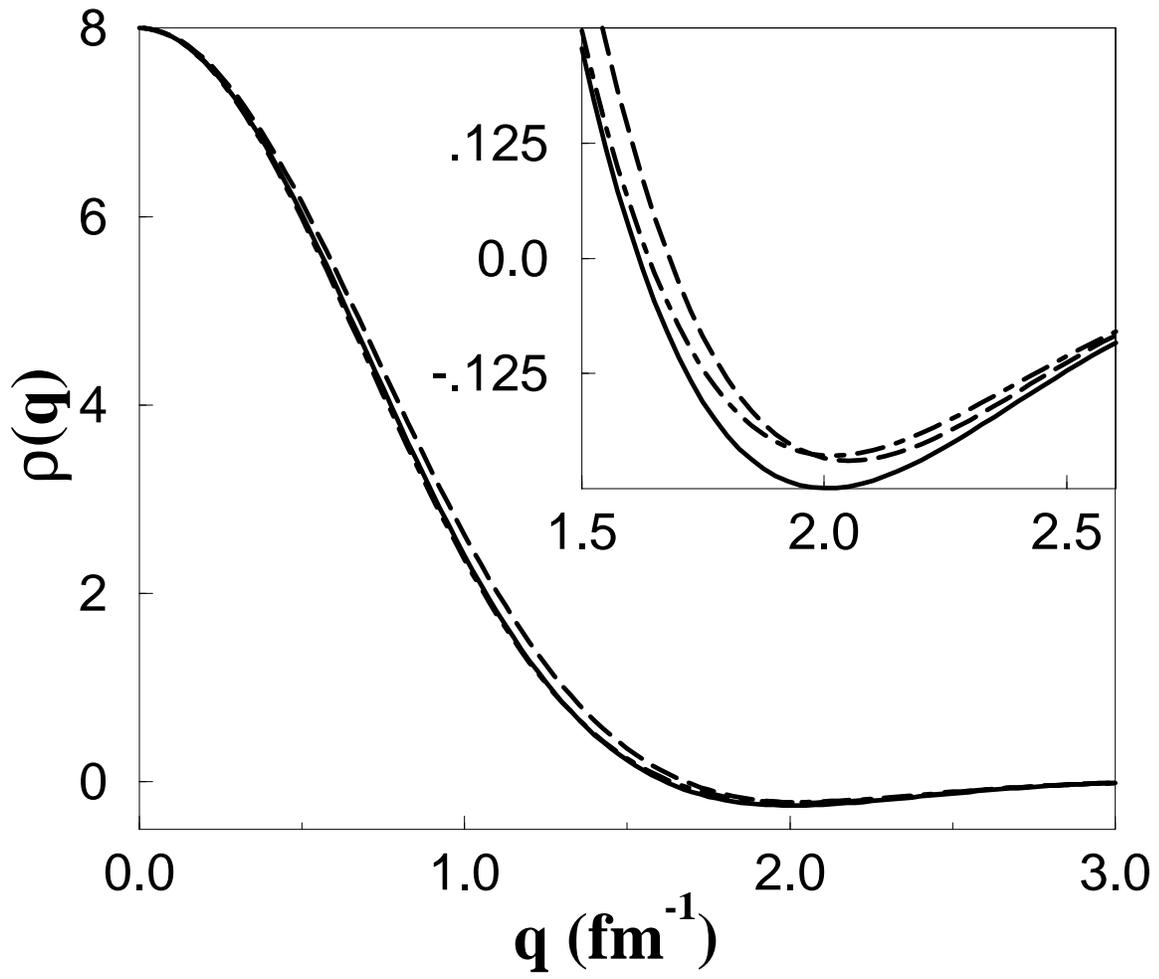
\begin{figure}
\caption{Comparison of the experimental proton density profile $\rho_p(q)$
for $^{16}$O \protect\cite{edata} (solid line)  with the calculated proton
density profiles from the DH model (dash-dotted line) and the HFB model
(dashed line). \label{fig2}}
\end{figure}

\noindent
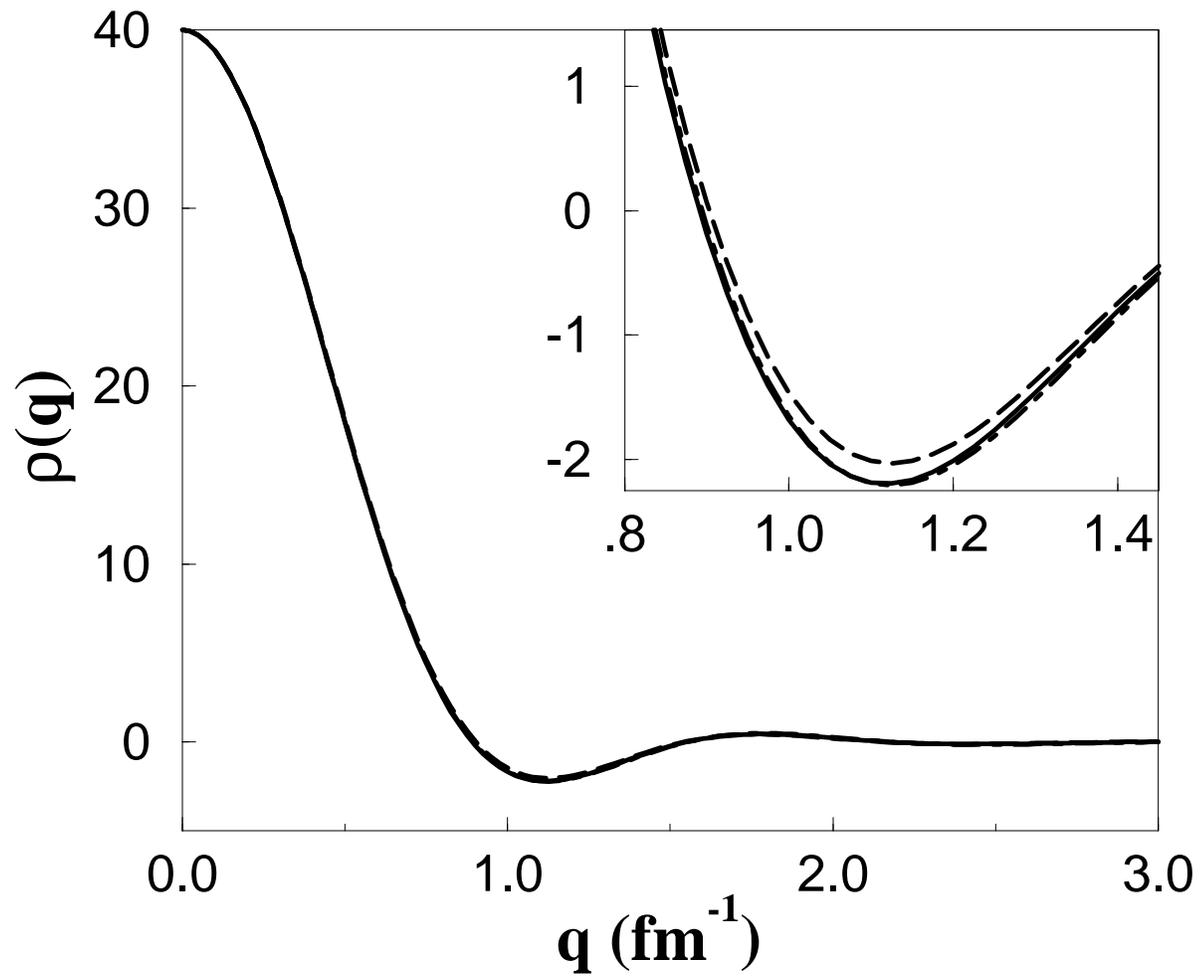
\begin{figure}
\caption{Comparison of the experimental proton density profile $\rho_p(q)$ 
for $^{90}$Zr \protect\cite{edata}  (solid line)  with the calculated proton 
density profiles from the DH model (dash-dotted line) and the HFB model 
(dashed line). \label{fig3}}
\end{figure}

\noindent
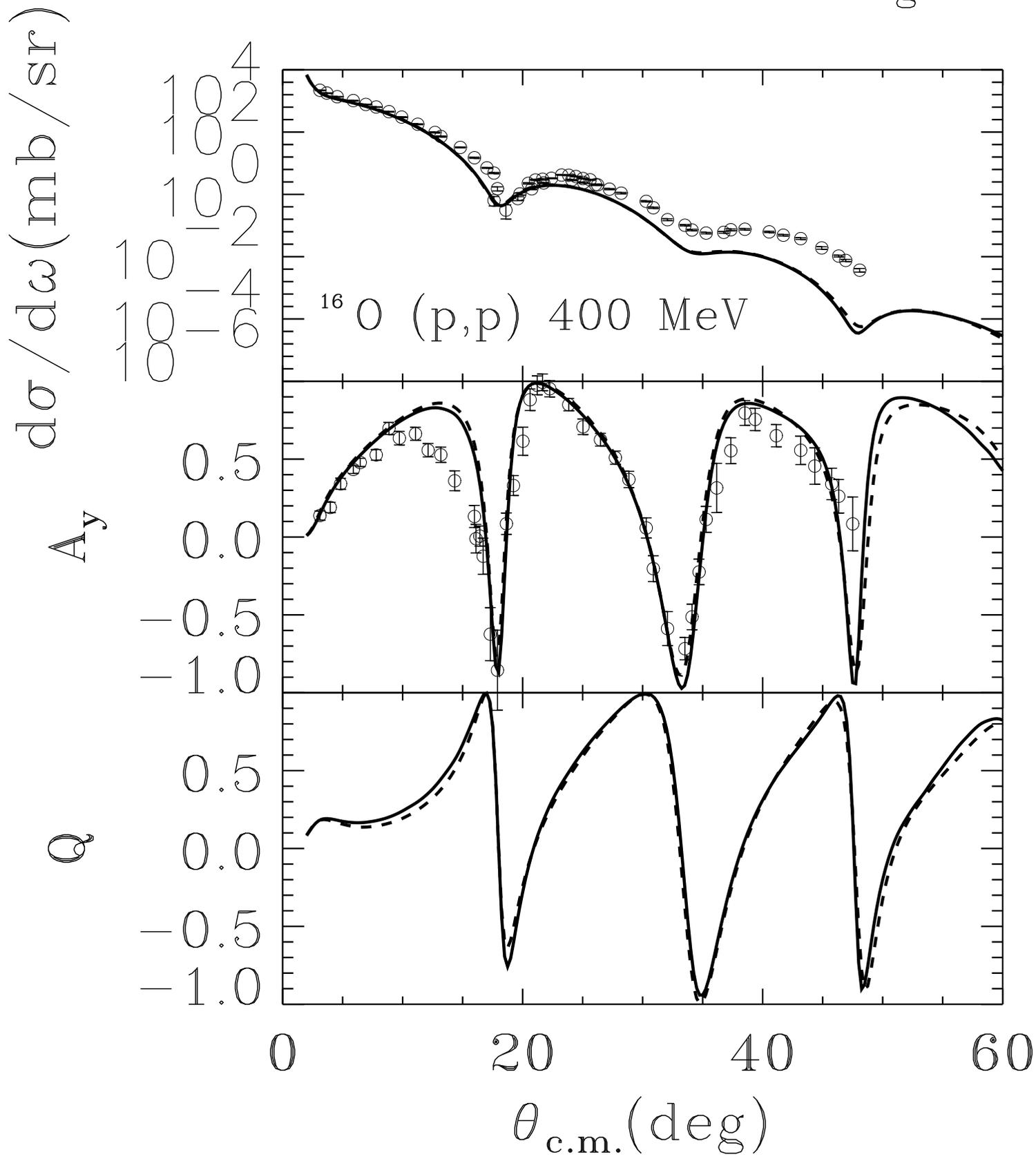
\begin{figure}
\caption{The angular distribution of the differential cross-section
         ($\frac{d\sigma}{d\Omega }$), analyzing power ($A_y$) and
         spin rotation function ($Q$) are shown for elastic proton
         scattering from $^{16}$O at 400 MeV laboratory energy.
         The solid line represents the calculation performed with a
first-order full-folding optical potential based on the DH density
\protect\cite{DH} and the Bonn model D52 \protect\cite{D52}. The dashed
line represent the calculation using the factorized, 
off-shell `$t\rho$' approximation to this optical potential.
The data are taken from Ref.~\protect\cite{O400}.\label{fig4} }
\end{figure}

\noindent
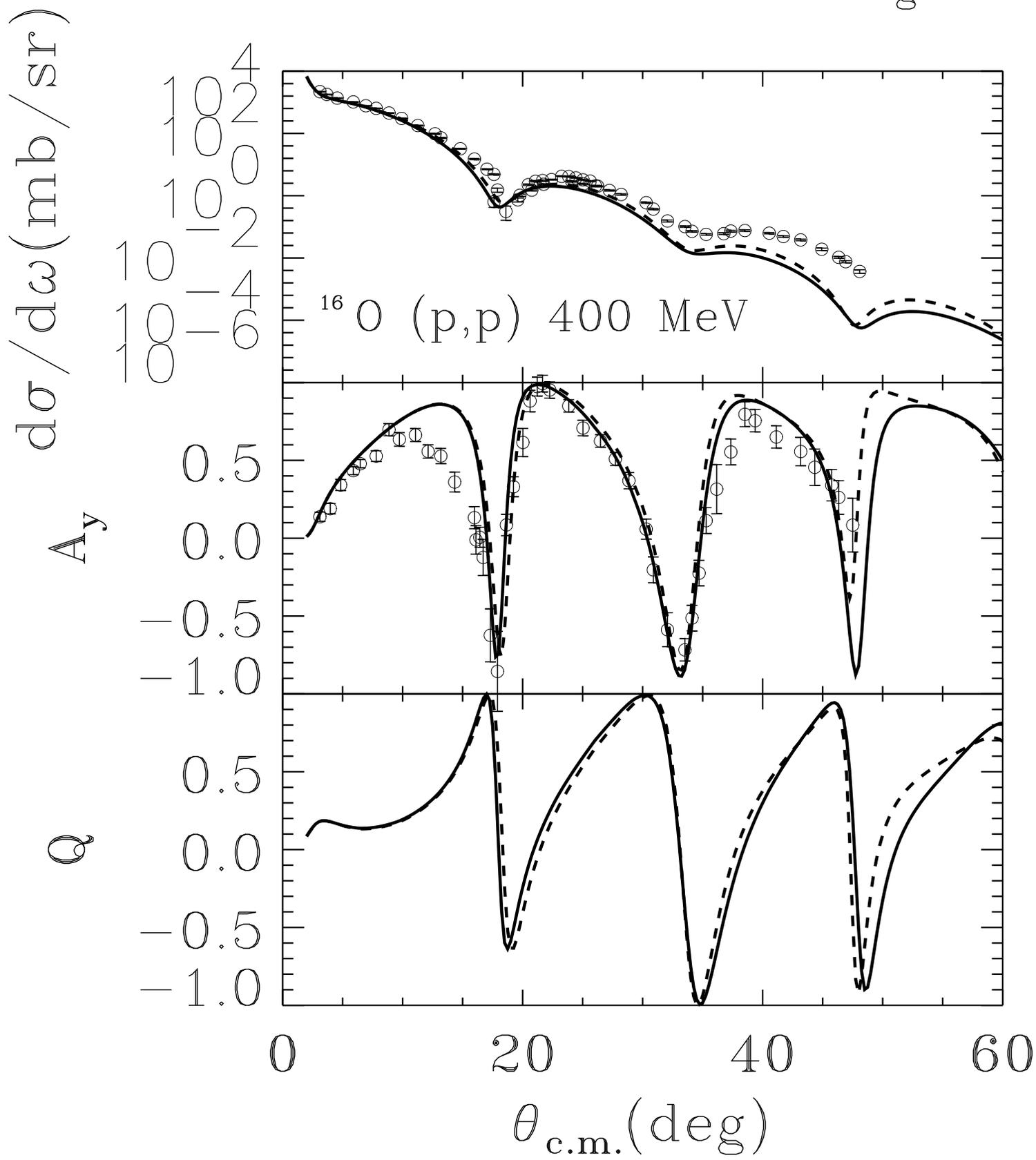
\begin{figure}
\caption{Same as Fig.~4, except that all calculations are based on the 
factorized, off-shell `$t\rho$' approximation. The solid line represents
the calculation using the DH \protect\cite{DH} density profile, whereas
the dashed line uses the HFB \protect\cite{HFB} density profile. 
\label{fig5} }
\end{figure}

\noindent
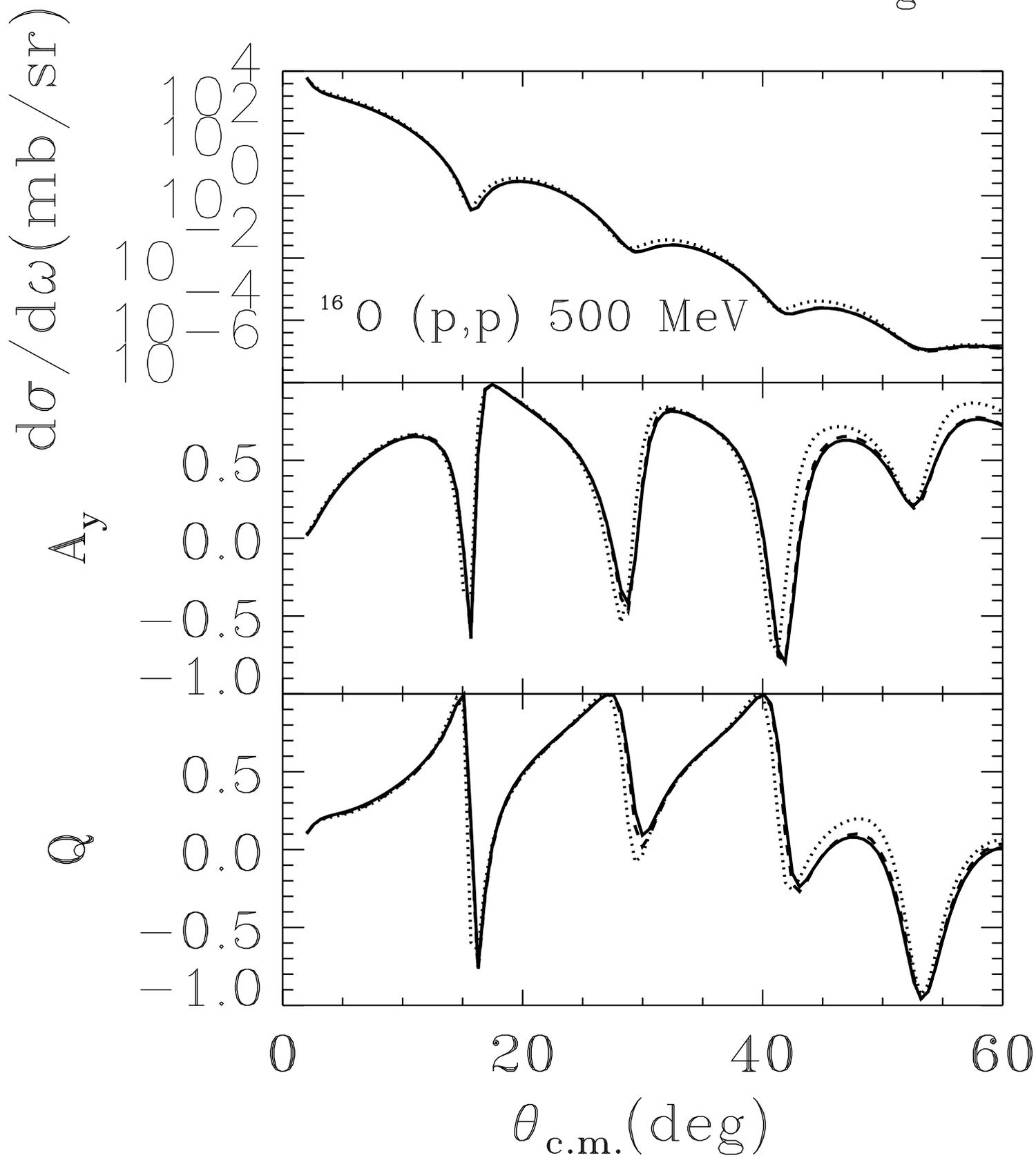
\begin{figure}
\caption{The angular distribution of the differential cross-section
         ($\frac{d\sigma}{d\Omega }$), analyzing power ($A_y$) and
         spin rotation function ($Q$) are shown for elastic proton
         scattering from $^{16}$O at 500 MeV laboratory energy.
The solid line represent the calculation performed with a first-order
full-folding optical potential as described in Section II. The dashed
line represents a calculation where the M\o ller factor is 
evaluated for the fixed angle $\Theta=90^o$, whereas for the
dotted line the M\o ller factor was omitted altogether.
All calculations are based on the DH density and the Bonn model D52 
t-matrix. \label{fig6} }
\end{figure}
 
\noindent
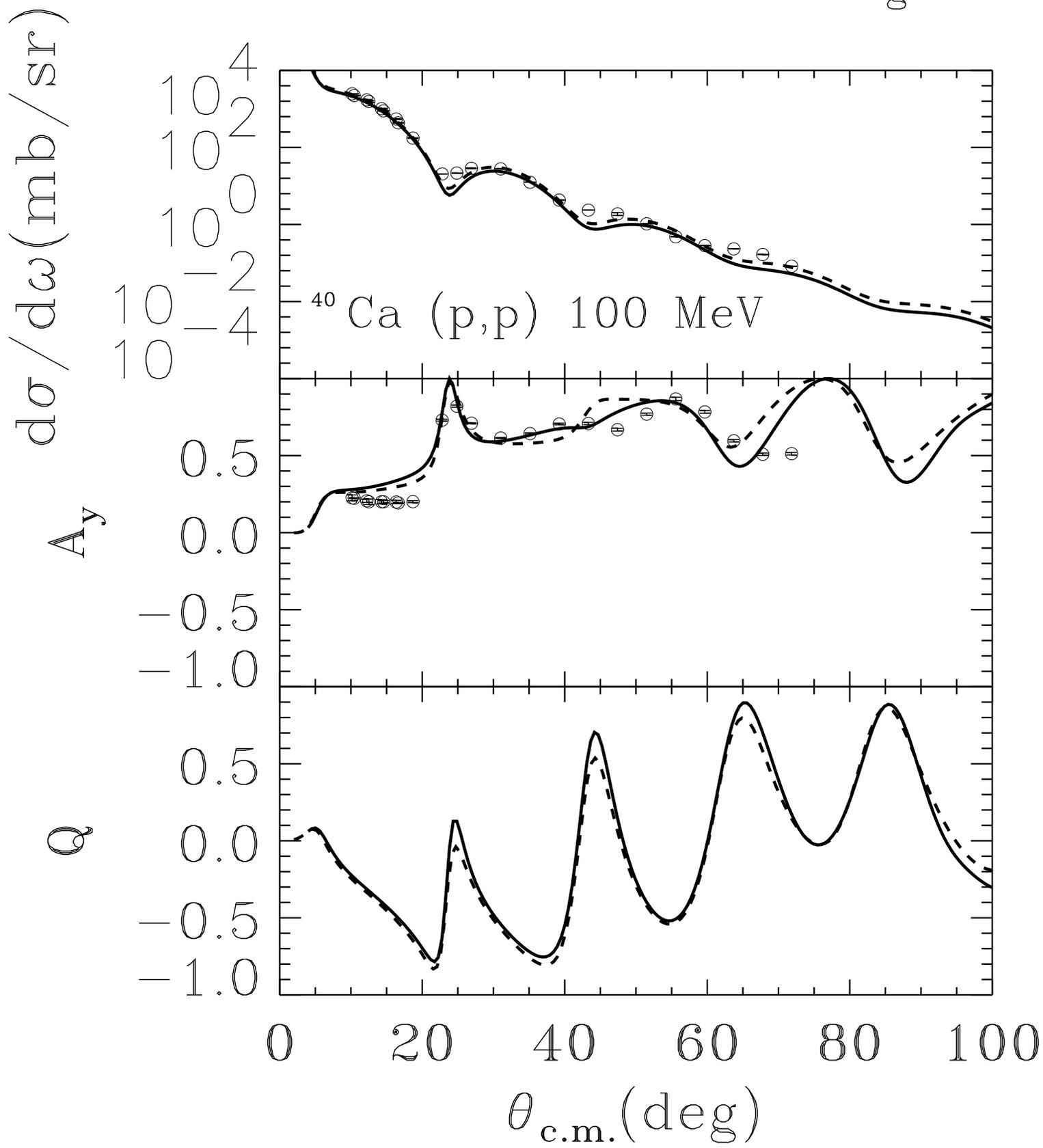
\begin{figure}
\caption{ The angular distribution of the differential cross-section
         ($\frac{d\sigma}{d\Omega }$), analyzing power ($A_y$) and
         spin rotation function ($Q$) are shown for elastic proton
         scattering from $^{40}$Ca at 100 MeV laboratory energy.
The solid line represent the calculation performed with a first-order
full-folding optical potential based on the DH density \protect\cite{DH}
and the full Bonn model \protect\cite{Bonn}, the dashed curve is based
on the factorized, off-shell `$t\rho$' approximations.
The data are taken from
Ref.~\protect\cite{ca100}.\label{fig7} }
\end{figure}

\noindent
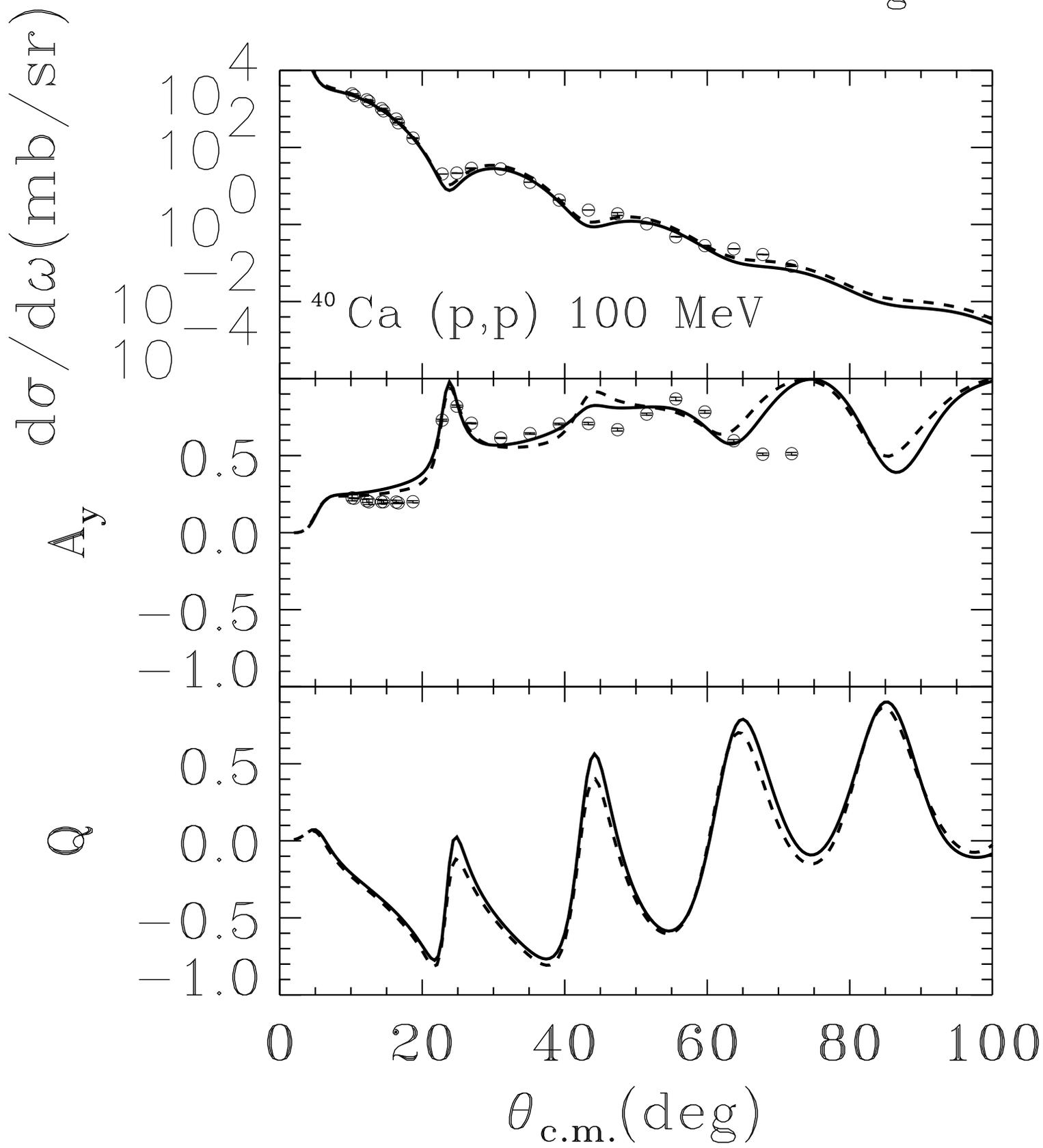
\begin{figure}
\caption{ Same as Fig.~7, except that the HFB model 
\protect\cite{HFB} is employed for
the density as well as the mean field force.  \label{fig8} }
\end{figure} 

\noindent
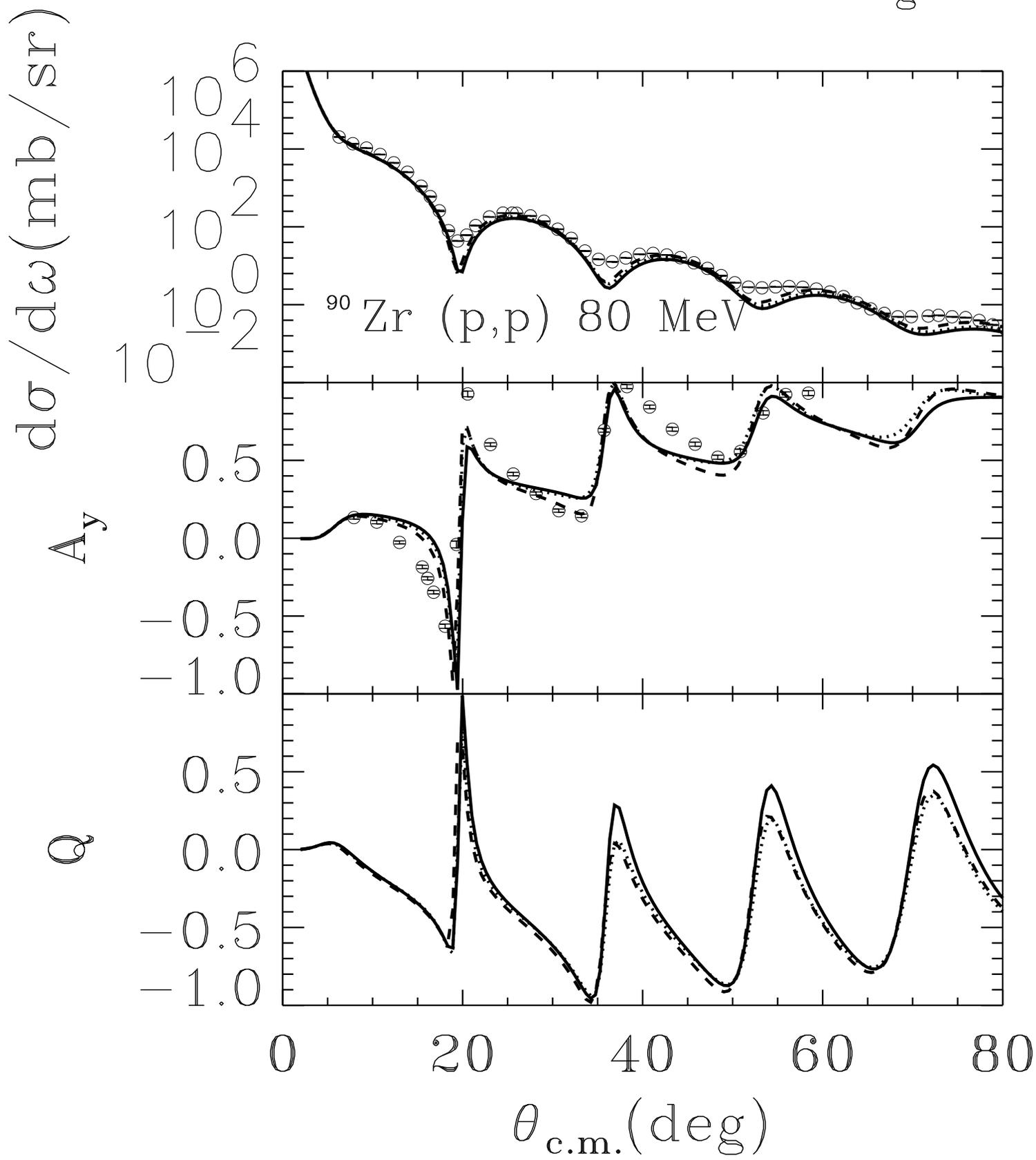
\begin{figure}
\caption{The angular distribution of the differential cross-section
         ($\frac{d\sigma}{d\Omega }$), analyzing power ($A_y$) and
         spin rotation function ($Q$) are shown for elastic proton
         scattering from $^{90}$Zr at 80 MeV laboratory energy.
The solid line represent the calculation performed with a first-order
full-folding optical potential based on the DH density \protect\cite{DH}
and the full Bonn model \protect\cite{Bonn}, the dashed curve is based
on the factorized, off-shell `$t\rho$' approximations.
The dotted line represents a full-folding calculations based on the
HFB model \protect\cite{HFB}. The data are taken from
Ref.~\protect\cite{zr80}. \label{fig9} }
\end{figure}

\noindent
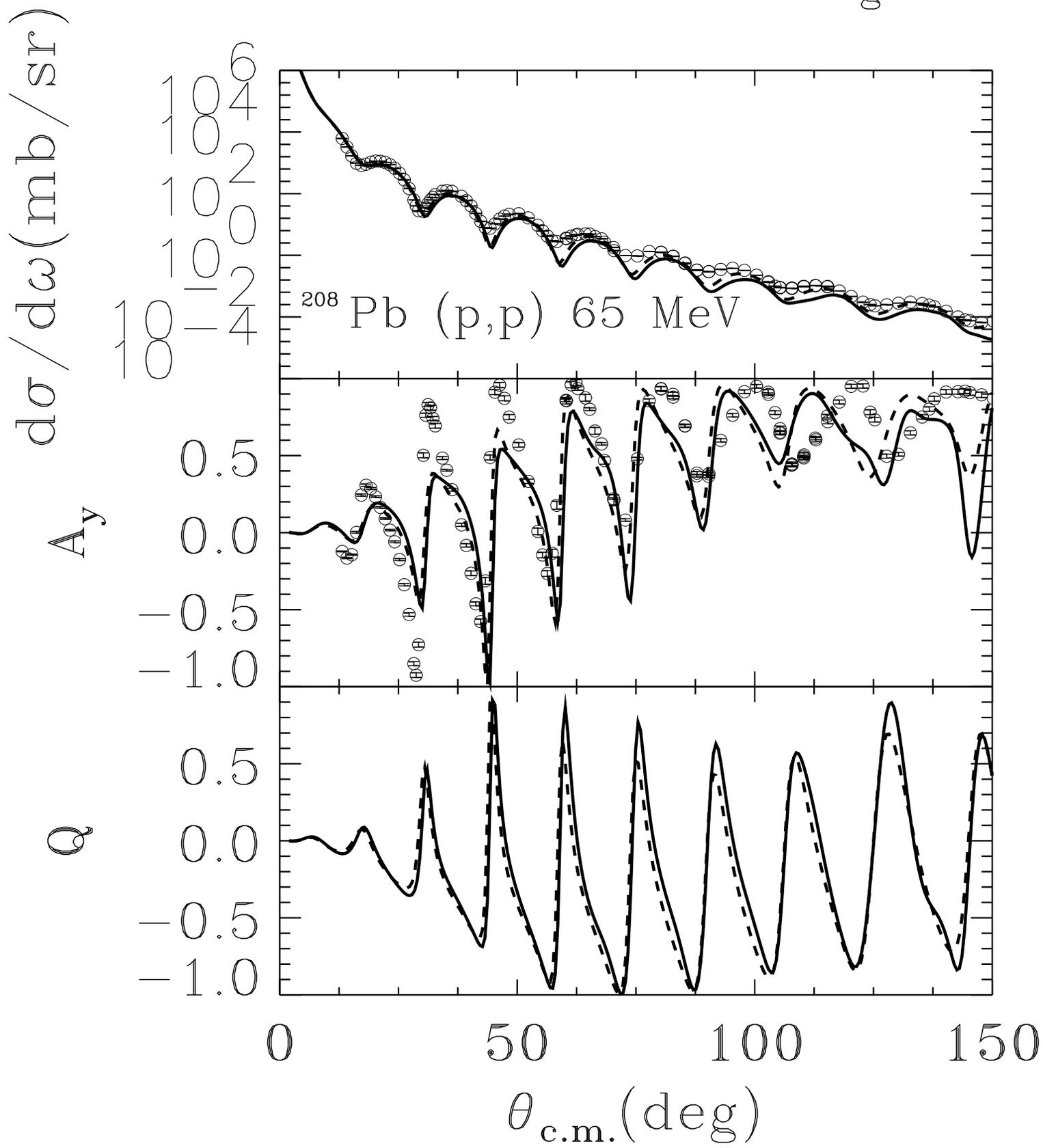
\begin{figure}
\caption{The angular distribution of the differential cross-section
         ($\frac{d\sigma}{d\Omega }$), analyzing power ($A_y$) and
         spin rotation function ($Q$) are shown for elastic proton
         scattering from $^{208}$Pb at 65 MeV laboratory energy.
The solid line represent the calculation performed with a first-order
full-folding optical potential based on the DH density \protect\cite{DH}
and the full Bonn model \protect\cite{Bonn}, the dashed curve is based
on the factorized, off-shell `$t\rho$' approximations.
The data are taken from
Ref.~\protect\cite{zr65}. \label{fig10} }
\end{figure}

\noindent
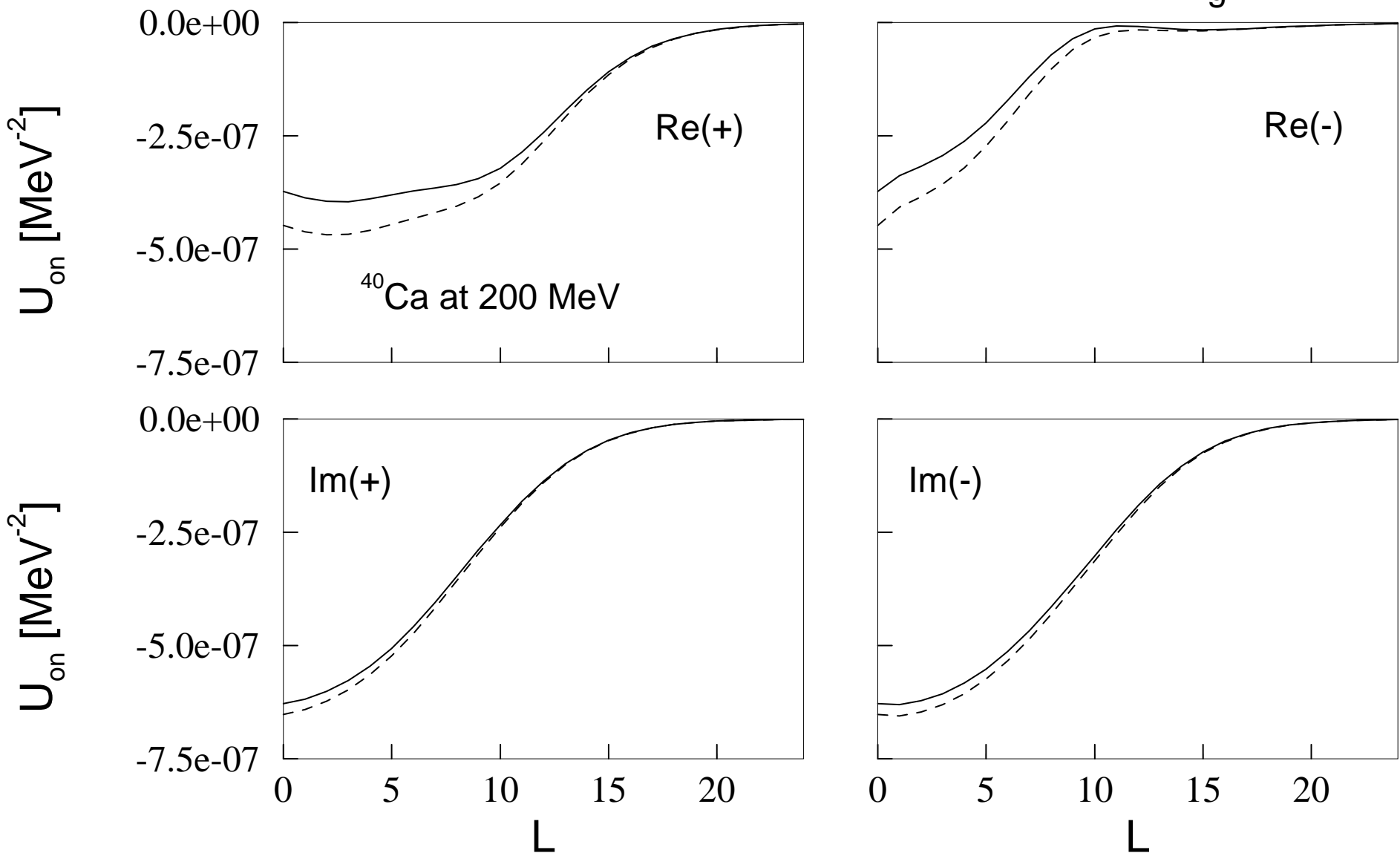
\begin{figure}
\caption{ Real and imaginary part of the on-shell value of the optical
potential as function of the orbital angular momentum L for scattering
from $^{40}$Ca at 200 MeV laboratory energy. (+) denotes the potential
for $J=L+\frac {1}{2}$, whereas (-) stand for $J=L-\frac {1}{2}$. The
full-folding (solid) and factorized, off-shell (dashed) calculations are
based on the full Bonn model and the DH density. \label{fig11} }
\end{figure}

\noindent 
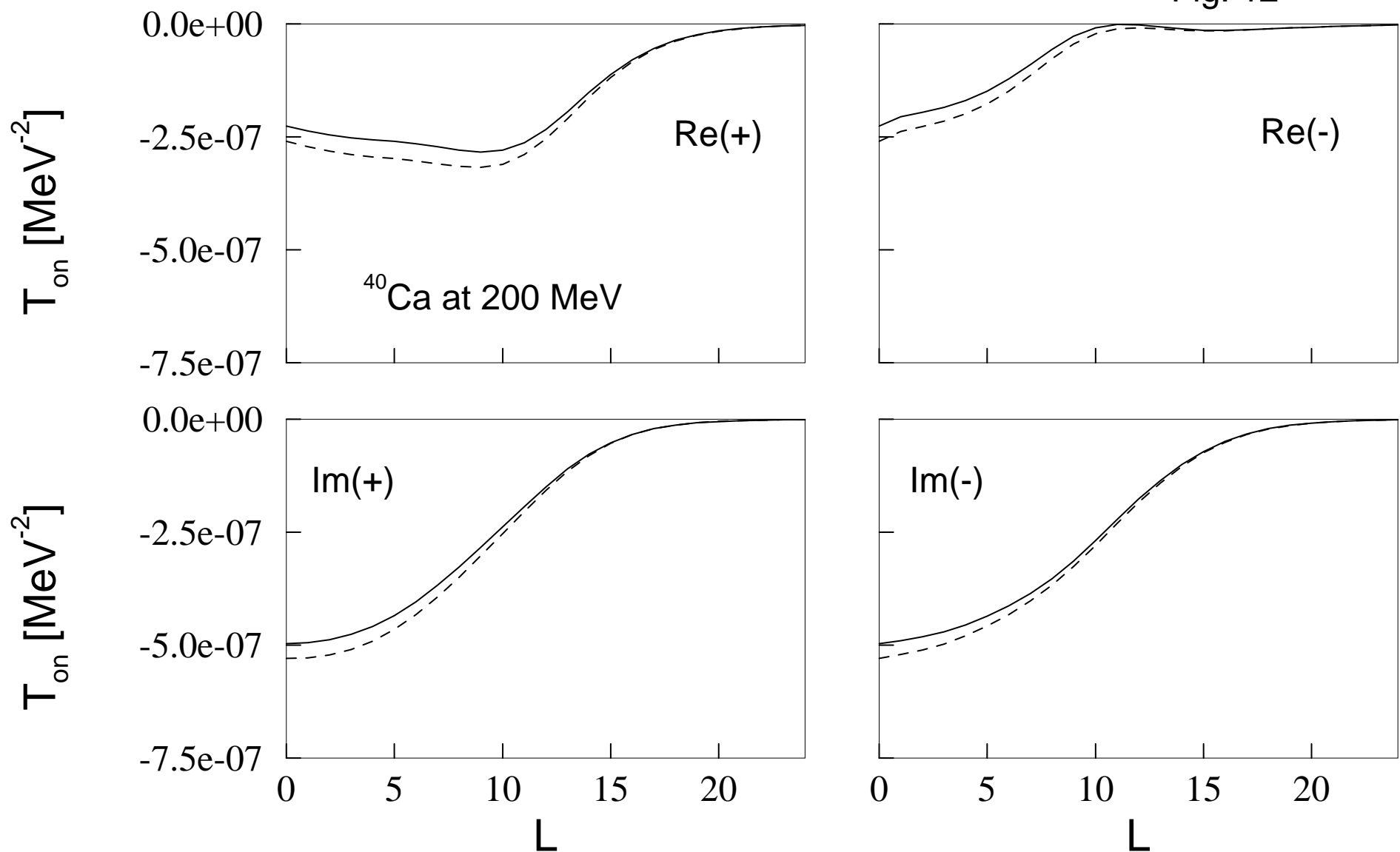
\begin{figure}
\caption{ Real and imaginary part of the on-shell value of the proton-nucleus
t-matrix (coulomb contributions omitted) as function of the orbital
angular momentum L for scattering
from $^{40}$Ca at 200 MeV laboratory energy. (+) denotes the potential
for $J=L+\frac {1}{2}$, whereas (-) stand for $J=L-\frac {1}{2}$. The
full-folding (solid) and factorized, off-shell (dashed) calculations are
based on the full Bonn model and the DH density. \label{fig12} }
\end{figure}

\noindent  
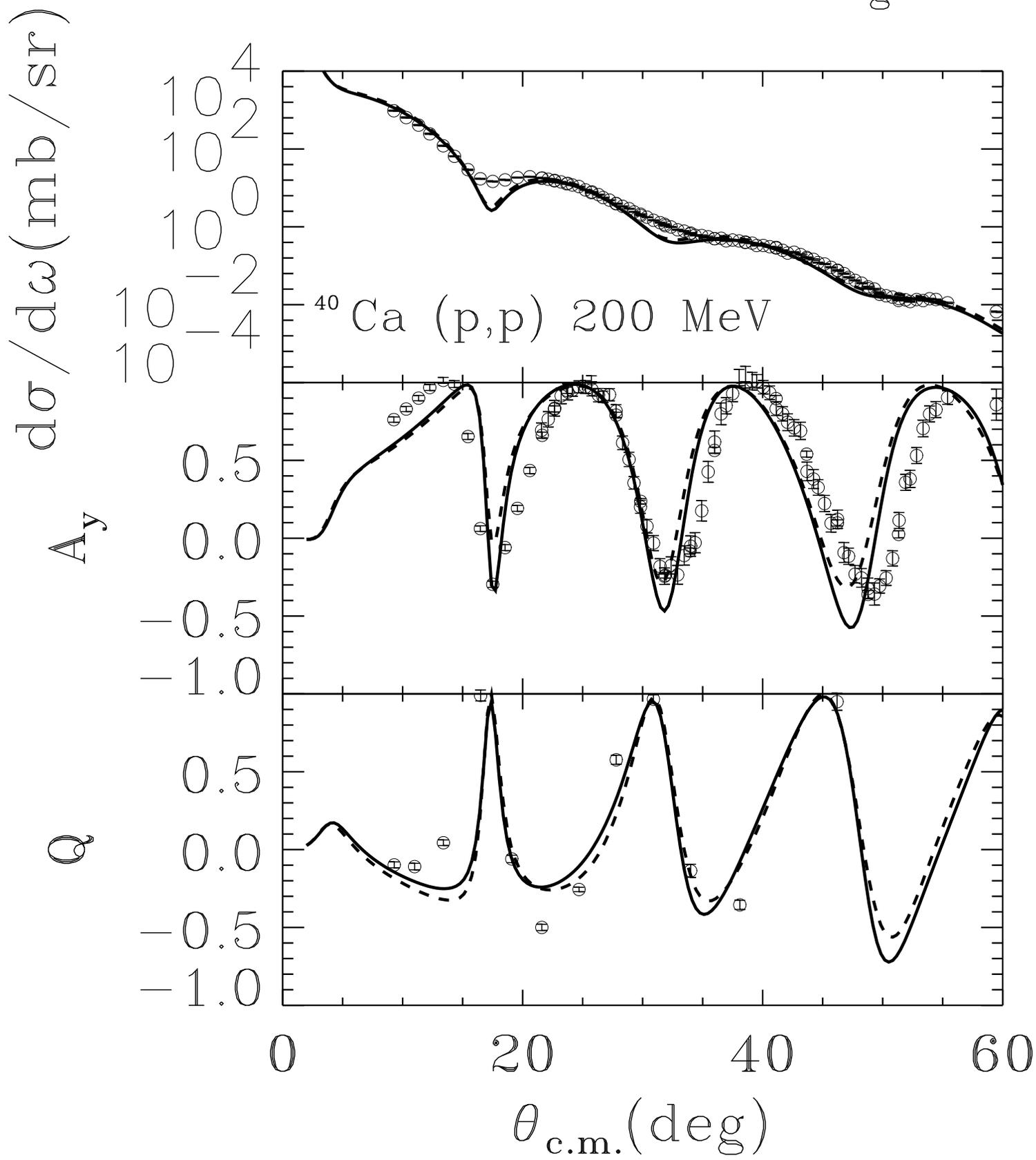
\begin{figure} 
\caption{The angular distribution of the differential cross-section
         ($\frac{d\sigma}{d\Omega}$), analyzing power ($A_y$) and
         spin rotation function ($Q$) are shown for elastic proton
         scattering from $^{40}$Ca at 200 MeV laboratory energy.
The solid line represent the calculation performed with a first-order
full-folding optical potential based on the DH density \protect\cite{DH}
and the full Bonn model \protect\cite{Bonn}, the dashed curve is based
on the factorized, off-shell `$t\rho$' approximation.
The data are taken from
Ref.~\protect\cite{O200}. \label{fig13} }
\end{figure}

\noindent
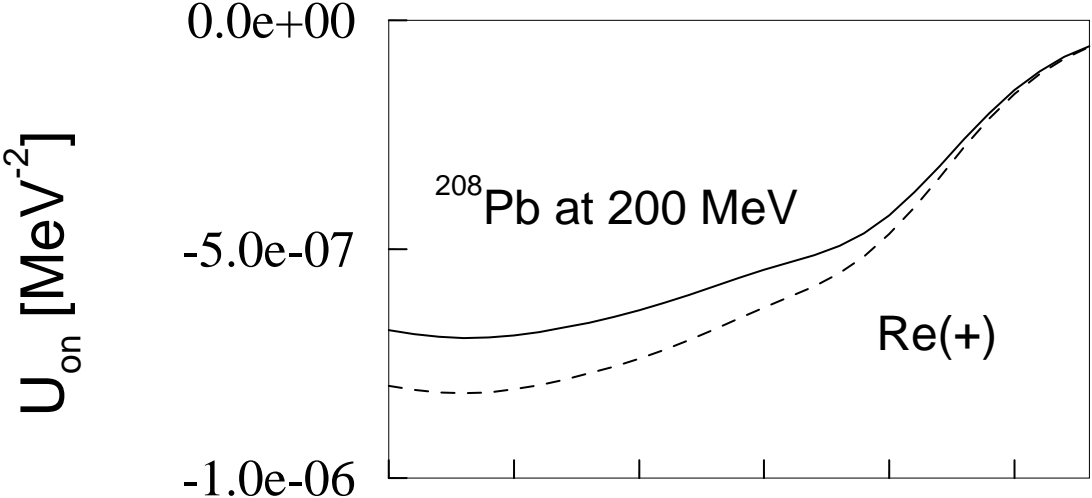
\begin{figure}
\caption{Same as Fig.~11, except for $^{208}$Pb. \label{fig14} }
\end{figure}
 
\noindent 
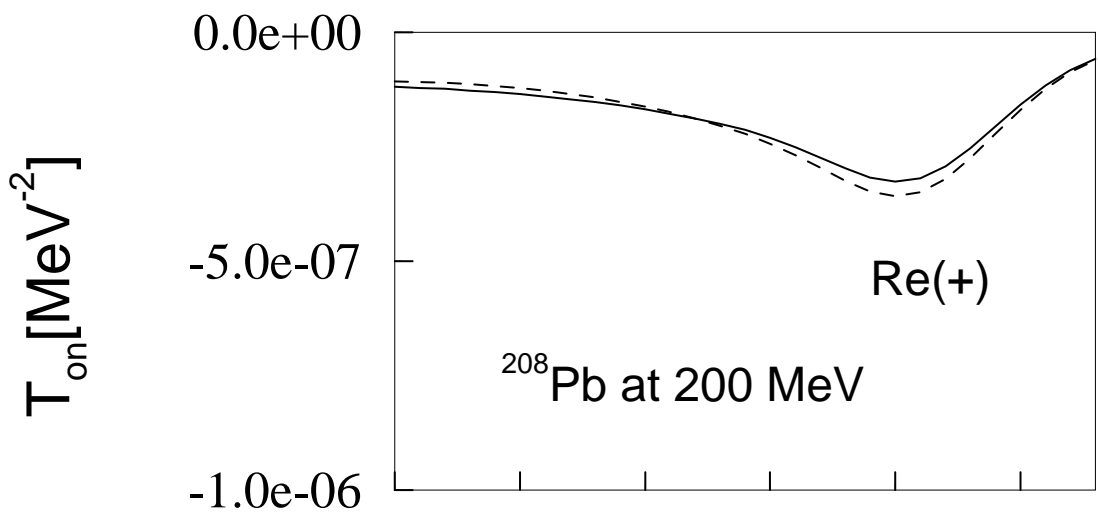
\begin{figure} 
\caption{Same as Fig.~12, except for $^{208}$Pb. \label{fig15} }
\end{figure}

\noindent 
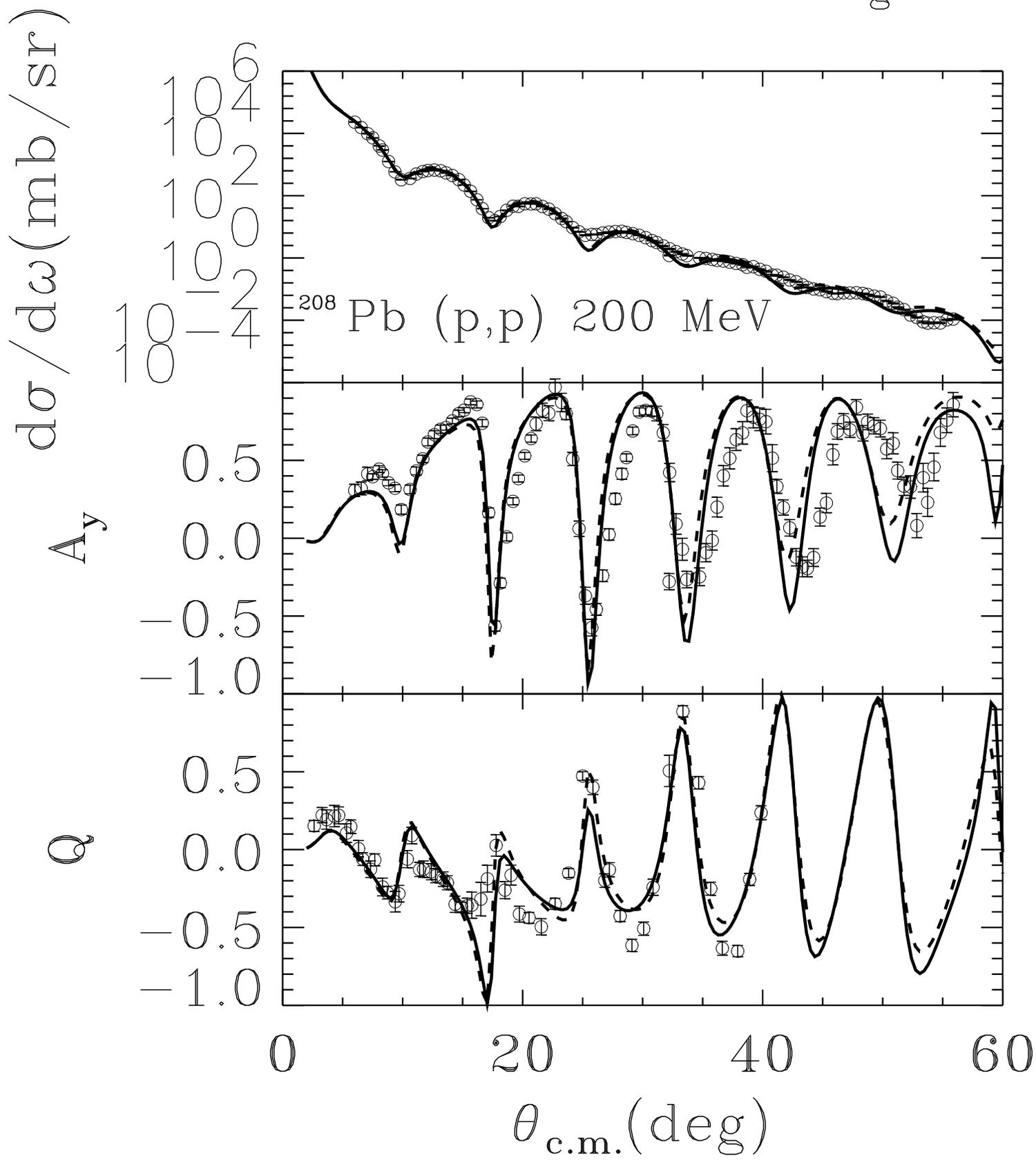
\begin{figure}  
\caption{Same as Fig.~13, except for $^{208}$Pb. The data are taken from
Ref.~\protect\cite{pb200}. \label{fig16} }
\end{figure}

\noindent
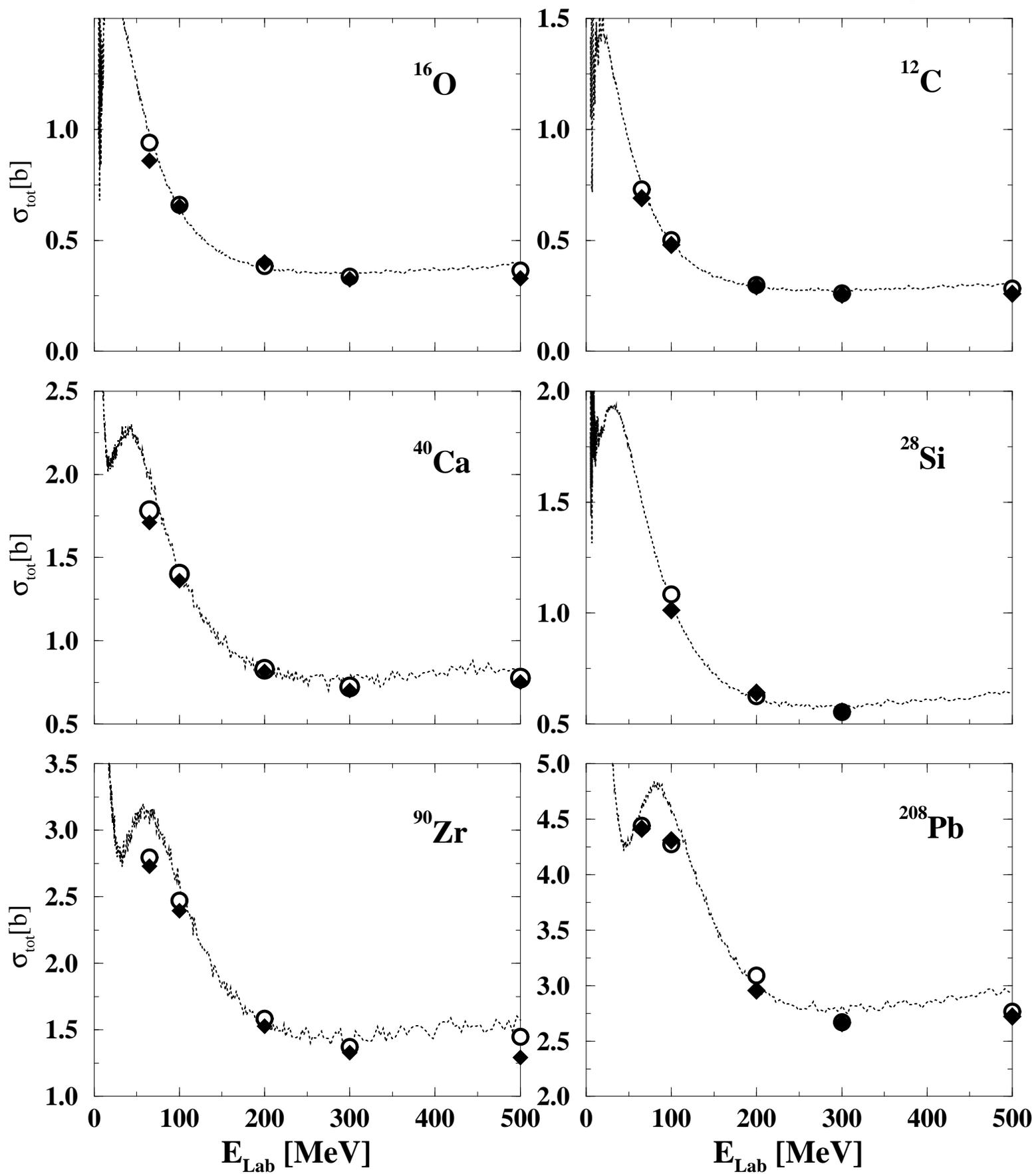
\begin{figure}
\caption{The neutron-nucleus  total cross-sections for
         scattering from $^{12}$C, $^{16}$O, $^{28}$Si, $^{40}$Ca,
         $^{90}$Zr, and $^{208}$Pb are shown as
         a function of the incident neutron kinetic energy.
         The dotted line represents the data taken from
         Ref.~\protect\cite{ndata,Roger}. The solid diamonds correspond
to the full-folding calculations using the full Bonn NN t-matrix
\protect\cite{Bonn} and the DH model \protect\cite{DH} for the density.
The open circles correspond to the factorized, off-shell `$t\rho$'
approximation. The calculations for energies smaller and equal 200 MeV
include the propagator modification due to the DH mean field.
\label{fig17} }
\end{figure}


\begin{references}

\bibitem{med2} C.R.~Chinn, Ch.~Elster, R.M.~Thaler, and S.P.~Weppner,
Phys. Rev. {\bf C52}, 1992 (1995).

\bibitem{FFC} R. Crespo, R.C. Johnson, and J.A. Tostevin, Phys. Rev.
  {\bf C41}, 2257 (1990).
 
\bibitem{FFO}  T. Cheon, Ch.~Elster, E.F. Redish, and P.C. Tandy,
 Phys. Rev. {\bf C41}, 841 (1990).

\bibitem{hugo1} H.F. Arellano, F.A. Brieva, M. Sander, H.V. von Geramb,
Phys. Rev. {\bf C54}, 2570 (1996).
 
\bibitem{hugo2}
H.F. Arellano, F.A. Brieva, and W.G. Love, Phys.
Rev. {\bf C52}, 301 (1995);
H.F. Arellano, F.A. Brieva, and W.G. Love, Phys.
Rev. {\bf C41}, 2188 (1990).

\bibitem{Corr} D.J.~Ernst, J.T.~Londergan, G.A.~Miller, and R.M.~Thaler,
Phys. Rev. {\bf C16}, 537 (1977).
 
\bibitem{Sicil} E.R. Siciliano and R.M. Thaler, Phys. Rev. {\bf C16},
                1322 (1977).
\bibitem{TT} P.C.~Tandy and R.M.~Thaler, Phys. Rev. {\bf C22}, 2321
(1980).

\bibitem{Watson} K.M. Watson, Phys. Rev. {\bf 89}, 575 (1953); N.C. Francis
    and K. M. Watson, {\it ibid.} {\bf 92}, 291 (1953).
 
\bibitem{KMT} A. Kerman, M. McManus, and R.M. Thaler, Ann. Phys. {\bf 8},
     551 (1959).

\bibitem{med1} C.R.~Chinn, Ch.~Elster, R.M~Thaler, Phys. Rev.
{\bf C48}, 2956 (1993)

\bibitem{bolle} P.C. Tandy, E.F. Redish, and D. Boll\'{e}, Phys. Rev.
{\bf C16}, 1924 (1977)

\bibitem{pttw} A. Picklesimer, P.C.~Tandy, R.M. Thaler, and
               D. H.~Wolfe, Phys. Rev. {\bf C 30}, 1861 (1984).

\bibitem{joachain} C.J. Joachain, `Quantum Collision Theory' 
(North-Holland Physics 1987) p.~387;
 C.~M\o ller, Kgl. Danske Videnskab. Selskab., Mat.-fys. Medd
{\bf 23}, 1 (1959).

\bibitem{D52} Ch. Elster and P.C. Tandy, Phys. Rev. {\bf C40} (1989), 881. 


\bibitem{DH} C.J.~Horowitz and B.D.~Serot, Nucl. Phys {\bf A368}, 503 (1981).

\bibitem{HFB}See for example J.F. Berger, M.~Girod, and D.~Gogny,
 Nucl. Phys. {\bf A502}, 85c (1989); J.P.~Delaroche, M.~Girod, J.~Libert
and I.~Deloncle, Phys. Lett. {\bf B232}, 145 (1989).

\bibitem{Gogny} J.~Decharg\'{e} and D. Gogny, Phys. Rev. {\bf C21}, 1568
(1980); J.F.~Berger, M.~Girod, and D.~Gogny, Comput. Phys.
    Commun. {\bf 63}, 365 (1991).
    

\bibitem{TIMORA} C.J.~Horowitz, D.P.~Murdoch, and B.D.~Serot, 
{\it Computational Nuclear Physics 1}, edited by K.~Langanke,
J.A.~Maruhn, and S.E.~Koonin (Springer-Verlag, Berlin, 1991), p. 129
({\it includes source code})

\bibitem{edata} H.~De Vries, C.W.~De Jager, and C. De Vries,
Atomic and Nuclear Data Tables {\bf 36}, 495(1987)

\bibitem{sobol} W.H.~Press and S.A.~Tenkolsky, Comp. in Phys. {\bf 76}(1989)

\bibitem{Bonn} R. Machleidt, K. Holinde, and Ch. Elster, Phys. Rep.
{\bf 149}, (1987).

\bibitem{coul}C.R.~Chinn, Ch.~Elster, and R.M.~Thaler, Phys. Rev.
 {\bf C44}, 1569 (1991).
 
\bibitem{ernst} D.J.~Ernst and G.A.~Miller, Phys. Rev. {\bf C21}, 1472 (1980);
D.L.~Weiss and D.J.~Ernst, Phys. Rev. {\bf C26}, 605 (1982);
D.J.~Ernst, G.A.~Miller, and D.L.~Weiss, Phys. Rev. {\bf C27}, 2733 (1983)

\bibitem{electron} I.~Sick, J.~B.~Bellicard, J.~M.~Cavedon, B.~Frois,
                    M.~Huet, P.~Leconte, P.~X.~Ho, and S.~Platchkov,
                    Phys. Lett. {\bf 88B}, 245 (1979);
 B.~Frois, J.B.~Bellicard, J.M.~Cavedon, M.~Huet,
  P.~Leconte, P.~.Ludeau, A.~Nakada, and P.X.~Ho, Phys. Rev. Lett.
  {\bf 38}, 152 (1977).

\bibitem{ISO} C.R.~Chinn, Ch.~Elster, and R.M.~Thaler, Phys. Rev.
  {\bf C47}, 2242 (1993).

\bibitem{coul}C.R.~Chinn, Ch.~Elster, and R.M.~Thaler, Phys. Rev.
 {\bf C44}, 1569 (1991). 

\bibitem{ndata}  R. W. Finlay, W. P. Abfalterer, G.~Fink, E.~Montei,
                 T~Adami, P.~W.~Lisowski, G.~L.~Morgan and R.~C.~Haight,
                Phys. Rev.  {\bf C 47}, 237 (1993).
 
\bibitem{Roger} R.W.~Finlay, G.~Fink, W.~Abfalterer, P.~Lisowski,
G.L.~Morgan, and R.C.~Haight, in {\it Proceedings of the Internat.
Conference on Nuclear Data for
Science and Technology}, edited by S.M.~Qaim (Springer-Verlag, Berlin,
1992), p. 702.

\bibitem{O400} C.~Chan, M.S. thesis, University of Alberta, 1985;
 D.~Hutcheon, private communication.

\bibitem{ca100} H.~Seifert, Ph.D thesis, Univ. Maryland (1990).

\bibitem{zr80} P.~Schwandt, H.O.~Mayer, W.W.~Jacobs, A.D.~Baches,
S.E.~Vigdor, M.D.~Kartchuck, Phys. Rev. {\bf C26}, 55 (1982).

\bibitem{zr65} H.~Sakaguchi, M.~Nakamura, K.~Hatanaka, A.~Goto, T.~Noro,
              F.~Ohtani, H.~Sakamoto, H.~Ogawa, and S.~Kobayashi,
               Phys. Rev.  {\bf C 26}, 944 (1982).

\bibitem{O200} E.J.~Stephenson, in `Antinucleon- \& Nucleon-Nucleus
Interactions' Telluride, Co. 1985, pp 299, ed. by G.~Walker et al.
(Plenum Press, NY, 1985).

\bibitem{pb200} D.A.~Hutcheon et al. in `Polarizion Phenomena in Nuclear
Physics - 1980', Proceedings of the Fifth International Symposium on
Polarization Phenomena in Nuclear Physics, AIP Conf. Proc. No. 69,
edited by G.G.~Ohlson, R.E.~Brown, N.~Jarmie, W.W.~McNaughton, and
G.M.~Hale (AIP, NY, 1981), p. 454. The data for $Q$ are from Na Gi,
Masters Thesis, Simon Frazer University, British Columbia, 1987. 

\end{references}
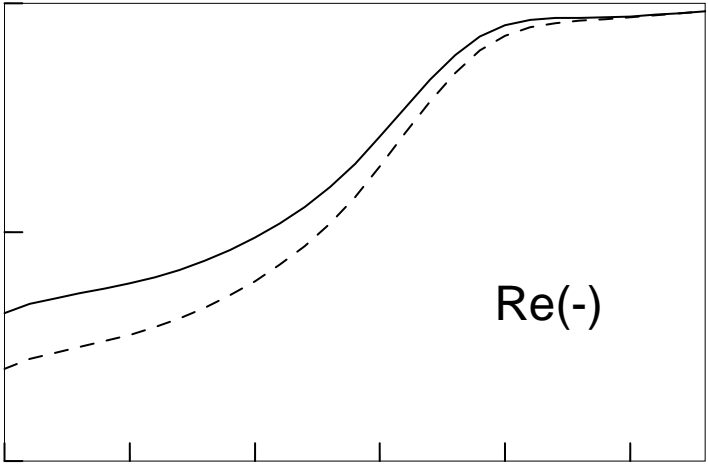
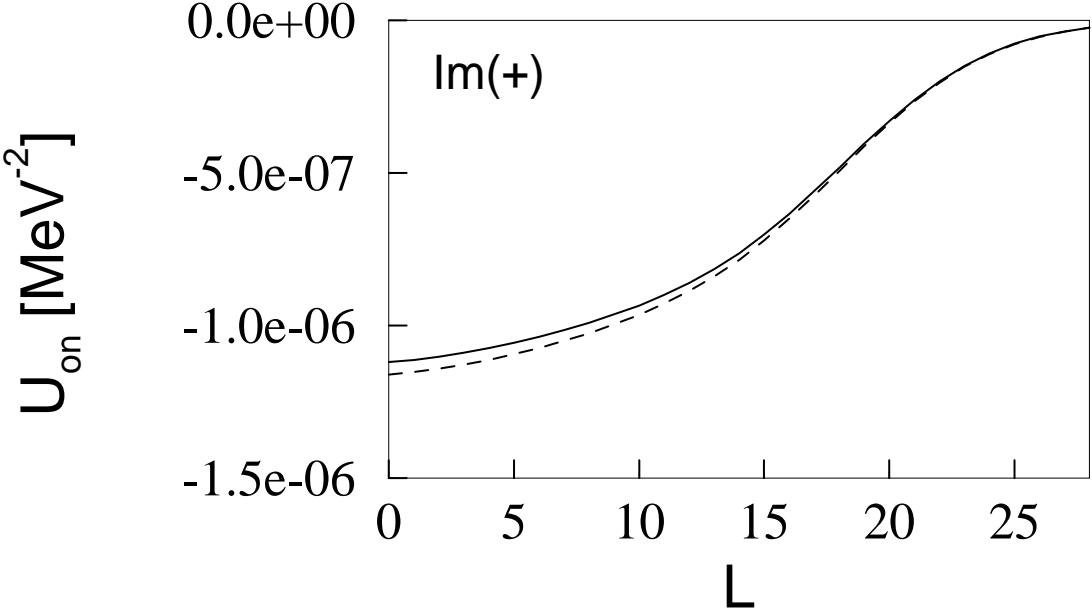
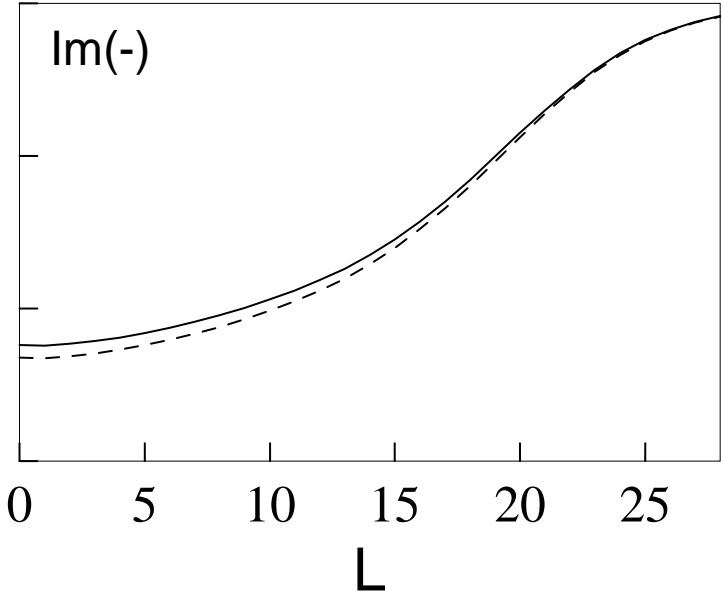
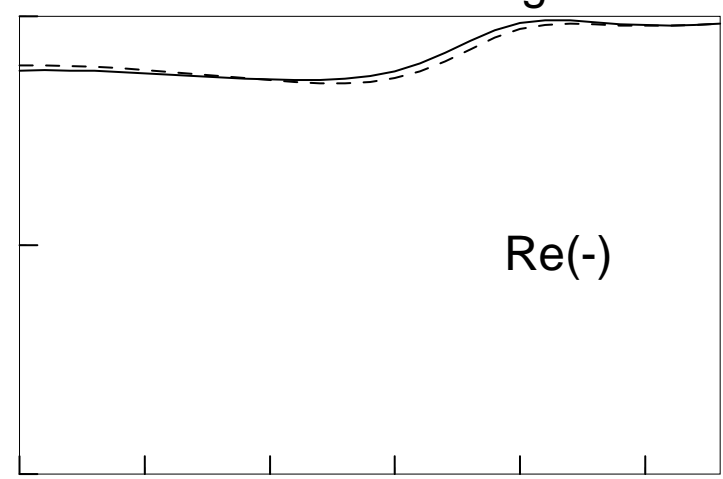
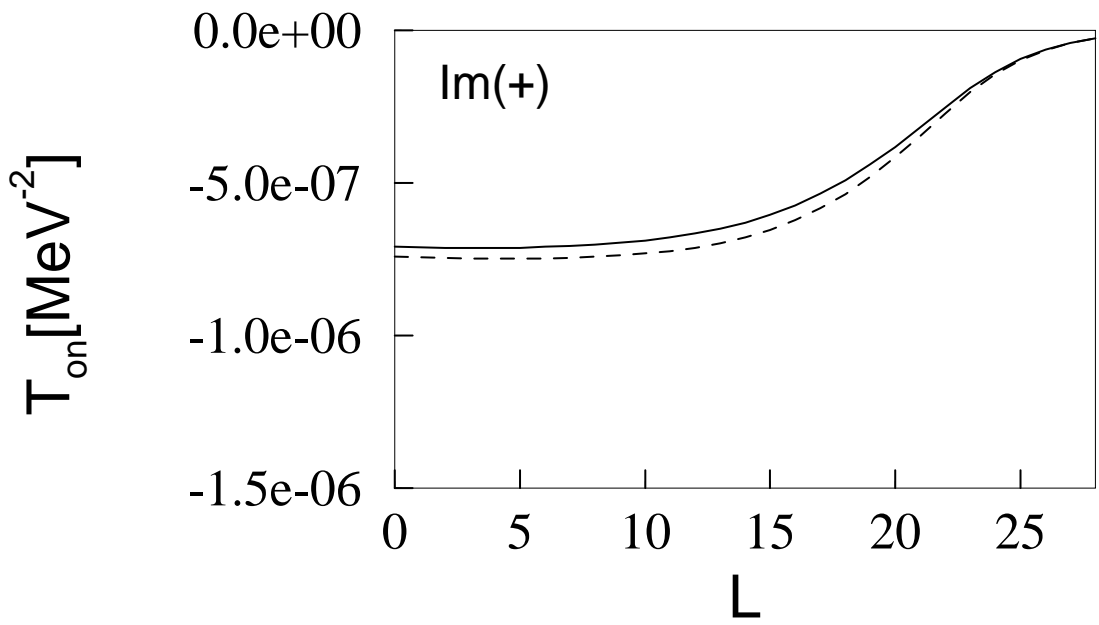
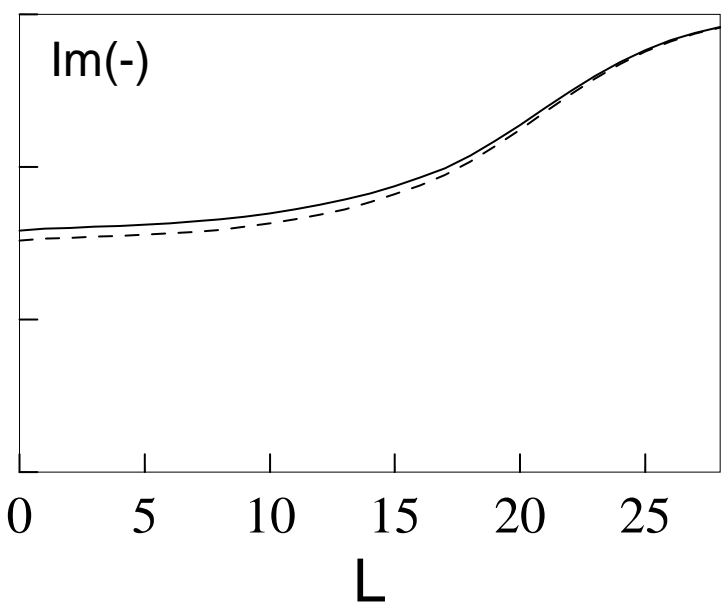
\end{document}